\documentclass[prb,reprint,superscriptaddress,twocolumn]{revtex4-2}
\usepackage{amsmath,amssymb,bm,braket,mathrsfs,mathtools}
\usepackage{multirow}
\usepackage{hyperref}

\usepackage[dvipdfmx]{graphicx}
\usepackage{color}

\DeclareMathOperator{\tr}{Tr}

\DeclareMathOperator{\diag}{diag}

\begin{document}

\allowdisplaybreaks

\title{Crystal-symmetry-protected gapless vortex-line phases in superconducting Dirac semimetals
}
\author {Shingo Kobayashi}
\affiliation{RIKEN Center for Emergent Matter Science, Wako, Saitama, 351-0198, Japan}
\author {Shuntaro Sumita}
\affiliation{Condensed Matter Theory Laboratory, RIKEN, Wako, Saitama, 351-0198, Japan}
\affiliation{Department of Basic Science, The University of Tokyo, Meguro-ku, Tokyo 153-8902, Japan}
\author {Motoaki Hirayama}
\affiliation{RIKEN Center for Emergent Matter Science, Wako, Saitama, 351-0198, Japan}
\affiliation{Department of Applied Physics, The University of Tokyo, Bunkyo-ku, Tokyo, 113-8656, Japan}
\author{Akira Furusaki}
\affiliation{RIKEN Center for Emergent Matter Science, Wako, Saitama, 351-0198, Japan}
\affiliation{Condensed Matter Theory Laboratory, RIKEN, Wako, Saitama, 351-0198, Japan}


\begin{abstract}
Vortex lines in superconducting Dirac semimetals realize crystal-symmetry-protected gapless vortex-line phases in which gapless excitations propagate inside a vortex line, in the presence of appropriate crystal symmetry, spin-orbit coupling, and multi-band structures.
Here we present a general scheme to classify possible gapless vortex-line phases in superconducting Dirac semimetals with rotation (or screw) symmetry and inversion symmetry, assuming that
the rotation (screw) axis is parallel to the vortex line.
The rotation (screw)-symmetry-protected gapless modes are stable as long as they have different rotation (screw) eigenvalues.  
The underlying mechanism for the formation of gapless vortex bound states depends on irreducible representations of rotation (screw) symmetry subject to a vortex field and is classified into three types: (i) accidental band crossing of two vortex bound-state modes under rotation symmetry; (ii) accidental and (iii) enforced band crossing of four vortex bound-state modes under screw symmetry.  
We present a tight-binding model of screw-symmetry-protected Dirac semimetal with an $s$-wave pair potential, demonstrating a gapless vortex-line phase of type (ii). We obtain four gapless modes of vortex bound states whose gapless points (Majorana zero modes) pinned at a time-reversal invariant momentum (TRIM) when the Fermi energy is close to the Dirac points.  As the Fermi energy is moved away from the Dirac points, the four gapless modes are split into a pair of two gapless modes with vanishing excitation energy at non-TRIMs.    
In closing, we discuss Nb$_3$Pt as a candidate material with the four-fold screw-symmetry-protected Dirac cones that can host a gapless vortex-line phase.
\end{abstract}
\maketitle

\section{Introduction}

Bound states in a vortex core of a conventional superconductor (SC) have energy spectrum with a mini gap of the order of $\Delta^2/E_F$, where $\Delta$ and $E_F$ are the superconducting gap and Fermi energy \cite{Caroli1964}.
The minigap vanishes and a Majorana zero mode is localized in a vortex core if the host superconductor is topological~\cite{Kopnin1991,Volovik1999fermion,ReadGreen2000,Volovik03,SatoFujimoto09,Ryu10,Teo10,SatoFujimoto16,Chiu16,Teo17review}.

A promising platform for Majorana zero modes is a hybrid system of a topological insulator (TI) and a conventional $s$-wave SC (SSC), where
a  Dirac cone on the surface of the three-dimensional (3D) TI is proximity-coupled to an $s$-wave pair potential,
yielding effectively a two-dimensional (2D) chiral $p$-wave state~\cite{Sato03,FuKane08}.
The theoretical proposal \cite{FuKane08} prompted experimental studies of TI/SSC hybrid systems~\cite{JPXu15,Sun16} and iron-based SCs~\cite{DWang18,PZhang18,QLiu18,Kong19,Machida19,kong2021majorana}.

A vortex line in a 3D SC is in a one-dimensional (1D) gapped topological SC phase of class D, when Majorana zero modes are localized at the two ends of the vortex line~\cite{Hosur11,GXu16,Qin19full}.
On the other hand, no Majorana zero mode can exist if the vortex line is in the topologically trivial phase of class D.
In both phases, vortex lines in 3D SCs have gapped bulk spectra of bound states.
Only when some parameter such as the Fermi energy is tuned to a topological phase transition point, a vortex line has a gapless spectrum of bound states \cite{Hosur11,Kobayash2020}.

This consideration leads to the following question:
Can a vortex line in a 3D SSC have a gapless phase, for a finite range of parameters, in which vortex bound states have a gapless excitation spectrum propagating along the vortex line?
A positive answer to this question has been given by theoretical studies of superconducting Weyl/Dirac semimetals~\cite{Volovik2011flat,Meng12,Konig2019,Qin2019,ZYan20,Giwa2021,Zhang2021}. 
As we discuss in detail in this paper,
superconducting 3D Weyl/Dirac semimetals can realize gapless vortex-line (GVL) phases in which vortex bound states form gapless modes traveling along a vortex line, the so-called 1D nodal vortex line \cite{Qin2019}.

The GVL phases are intrinsically related to bulk Dirac cones of the normal state via a low-energy effective Jackiw-Rossi model~\cite{JackiwRossi1981}, which is a model of 2D Dirac fermions coupled to a vortex field.
Suppose, for example, that the normal-state band structure has Weyl or Dirac points near the Fermi energy in addition to a metallic band with a Fermi surface.  When the metallic band turns into an SSC, Weyl/Dirac fermions proximity coupled to the $s$-wave pair potential
can be mapped approximately to the Fu-Kane model \cite{FuKane08} or the Jackiw-Rossi model; a vortex line would then have a gapless mode.

The GVL phases associated with Dirac points are protected from a gap-opening by crystalline symmetry~\cite{Qin2019}, similarly to Dirac semimetals whose Dirac points are protected by either rotation symmetry or screw symmetry together with inversion symmetry.
A large variety of 3D Dirac semimetals have been theoretically proposed~\cite{Young2012,Yang2014classification,YangMorimoto,Gao2016,Wieder2016Double,Watanabe2016,Armitage2018,Zhi-Ming2022} and experimentally observed: Na$_3$Bi~\cite{Liu2014discovery,Xu2015observation}, Cd$_2$As$_3$~\cite{Liu2014stable,Neupane2014observation,He2014Quantum,Borisenko2014,Liang2015ultrahigh,Crassee2018}, TlBiSSe~\cite{Novak2015}, ZrTe$_5$~\cite{Chen2015Magnetoinfrared,Chen2015Optical,Zheng2016Transport,Liu2016zeeman,Li2016chiral}, Li(Fe$_{1-x}$Co$_x$)As~\cite{Zhang2019multiple}, $\alpha$-Sn~\cite{Xu2017}, and CaAuAs\cite{Nakayama2020}.
Thus, 3D Dirac semimetals have the potential of realizing various GVL phases, but the comprehensive picture is still missing.

In this paper, we propose a general theoretical framework to classify GVL phases of a vortex line in SSCs that are protected under rotation or screw symmetry in addition to inversion symmetry, which are crystalline symmetries common to 3D Dirac semimetals. The screw symmetry is a nonsymmorphic symmetry that consists of a non-primitive lattice translation and a rotation operation, and here we will focus on screw symmetry with a half lattice translation so that Dirac points always exist at the Brillouin zone boundary ($k_z=\pi$).

From symmetry analysis of effective low-energy Hamiltonians describing vortex bound states, we determine the stability condition for GVL phases in terms of crystal symmetry in the presence of a vortex line.
We show that there are three types of mechanisms for the formation of gapless modes: (i) accidental band crossing (ABC) in a minimal $2 \times 2$ effective Hamiltonian with rotation symmetry; (ii) ABC and (iii) enforced band crossing (EBC) in a minimal $4 \times 4$ effective Hamiltonian with screw symmetry. Here, an ABC occurs within a certain parameter range, while an EBC occurs independently of parameter values. The GVL phases of type (i) appear under $n$-fold rotation symmetry ($n=3,4,6$) when vortex bound states forming gapless modes have different rotation eigenvalues, as discussed in previous works~\cite{Qin2019,Konig2019}. As for the GVL phases in screw-symmetric systems, we point out that the four- and six-fold screw symmetries protect a pair of gapless modes through (ii) an ABC and (iii) an EBC, respectively.

Furthermore, we discuss a four-fold screw-symmetry-protected GVL phase for a tight-binding model with space group symmetry $P4_2/mmc$, which has two Dirac points, at $Z=(0,0,\pi)$ and $A=(\pi,\pi,\pi)$ in the 3D Brillouin zone, protected by the four-fold screw symmetry along the $z$ axis and inversion symmetry. Implementing a vortex line parallel to the screw axis in the $s$-wave pair potential, we numerically study an evolution of vortex-line phases with the variation of the chemical potential. When the chemical potential is close to the Dirac points, four Majorana zero modes are pinned at $k_z=\pi$ since the low-energy Hamiltonian is equivalent to four copies of the Jackiw-Rossi model~\cite{JackiwRossi1981}. When the chemical potential is changed away from the Dirac cones, the four gapless modes are split into a pair of two gapless modes located away from $k_z=\pi$, each pair of which are protected by the four-fold screw-symmetry. As the chemical potential is changed further, two pairs of gapless modes annihilate at $k_z=0$ and open a gap.

We note that the zero-energy bound states at time-reversal invariant momenta (e.g., $k_z=\pi$) are Majorana zero modes inside a vortex line.
This is an interesting feature of superconducting Dirac semimetals with screw symmetry having Dirac points at $k_z=\pi$.
In contrast, zero-energy bound states at $k_z\ne0$ or $\pi$ are not Majorana fermions, as they are linear combination of $c_{k_z}^{}$ and $c_{-k_z}^\dagger$, where $c_{k_z}$ is the electron annihilation operator with momentum $k_z$.
This is usually the case with superconducting Weyl semimetals and Dirac semimetals without screw symmetry.

In addition, we propose Nb$_3$Pt as a candidate material for a superconducting Dirac semimetal that has four-fold screw-symmetry-protected Dirac cones at the $X$ point and its symmetry related points in the Brillouin zone.

This paper is organized as follows. In Secs.~\ref{sec:sym_Dirac} and~\ref{sec:sym_SC}, we introduce symmetry operations in the Dirac semimetals and those in the $s$-wave superconducting states with a vortex line. We construct an effective low-energy Hamiltonian in Sec.~\ref{sec:basis_principle} and develop a classification of GVL phases under rotation and screw symmetries in Sec.~\ref{sec:classification}. In Sec.~\ref{sec:model}, we model a tight-binding Hamiltonian on a tetragonal lattice with four-fold screw-symmetry-protected Dirac conesl and numerically demonstrate a four-fold screw-symmetry-protected GVL phase in the superconducting state. The application to Nb$_3$Pt is discussed in Sec.~\ref{sec:material}.  We summarize our results in Sec.~\ref{sec:model}. 

In the appendices, we show  topological classifications of GVL phases using the K-theoretical method in Appendix~\ref{app:k-th}, vortex zero-energy solutions in a low-energy Hamiltonian in Appendix~\ref{app:low-h}, and influence of vortex core positions on the 4-fold screw-symmetry-protected GVL mode in Appendix~\ref{app:VCP}.

\section{Symmetry analysis of GVL phases}
\label{sec:sym_analysis}

\subsection{Symmetry in Dirac semimetals}
\label{sec:sym_Dirac}
We begin with discussion of symmetry properties of 3D Dirac semimetals of non-magnetic materials that are invariant under time-reversal (TR) transformation $T$ ($T^2=-1$) and inversion transformation $I$ ($I^2=1$).
All energy bands are doubly degenerate due to the Kramers degeneracy enforced by the $TI$ symmetry, the combination of $T$ and $I$. Therefore, a Dirac point with four-fold degeneracy can be formed when two energy bands cross. However, such band-crossing points are generally unstable and gapped out by band-mixing perturbations that are present under the $TI$ symmetry. This gap-opening can be prevented by an additional crystal symmetry that is ubiquitous in solids, e.g., $n$-fold rotation symmetry ($n=2,3,4,6$) or $n$-fold screw symmetry ($n=2,4,6$). In the presence of one of these symmetries, the Dirac points are stable on rotation (screw) symmetric lines in the Brillouin zone when two crossing energy bands have different rotation (screw) eigenvalues.

The minimal Hamiltonian of a Dirac semimetal has the form
\begin{equation}
\hat{H}= \sum_{\bm{k}} \sum_{s,s',\sigma,\sigma'} c_{\bm{k}, s, \sigma}^{\dagger} H_{s,\sigma;s',\sigma'}(\bm{k})c_{\bm{k}, s', \sigma'},
\label{normal Hamiltonian}
\end{equation}
where $s$ and $s'$ are spin indices ($s \in \{\uparrow, \downarrow\}$), $\sigma$ and $\sigma'$ are orbital (sublattice) indices ($\sigma \in \{1, 2\}$). $ c_{\bm{k}, s, \sigma}^{\dagger}$ ($ c_{\bm{k}, s, \sigma}$) is the creation (annihilation) operator of an electron with wave number $\bm{k}$.
The action of TR, inversion, $n$-fold rotation, and screw transformations on the Hamiltonian in momentum space are defined by
\begin{subequations}\label{eq:dsm_sym}
\begin{align}
&TH(\bm{k})T^{-1} = H(-\bm{k}), \label{eq:trs} \\
&IH(\bm{k})I^{-1} = H(-\bm{k}),  \label{eq:is} \\
&C_n H(\bm{k}) C_n^{-1} = H(R_n\bm{k}), \label{eq:rs} \\
&S_n^{k_z} H(\bm{k}) (S_n^{k_z})^{-1} = H(R_n\bm{k}), \label{eq:ss}
\end{align}
\end{subequations}
where $R_n$ represents a rotation around the rotation (screw) axis, chosen to be the $z$ axis, in the momentum space:
\begin{align}
 R_n \bm{k} = \begin{pmatrix} \cos\left( \frac{2\pi}{n} \right) & -\sin \left( \frac{2\pi}{n} \right) & 0 \\ \sin\left(  \frac{2\pi}{n} \right) & \cos\left(  \frac{2\pi}{n} \right) & 0 \\ 0& 0 & 1 \end{pmatrix} 
                    \left(\begin{array}{@{\,}c@{\,}} k_x \\ k_y \\ k_z \end{array}\right). \label{eq:rotation}
\end{align}
The TR operator is given by $T=i\sigma_0 s_y K$, where $\sigma_0=s_0=\bm{1}_2$, $s_i$ ($i=x,y,z$) are the spin Pauli matrices, $K$ is the complex conjugation operator, and $\bm{1}_m$ is the $m \times m$ identity matrix. The unitary matrices $I$, $C_n$, and $S_n^{k_z}$ in Eqs.~(\ref{eq:dsm_sym}) represent inversion, rotation, and screw operators and satisfy the following relations:
\begin{subequations}\label{eq:dsm_sympro}
\begin{align}
&I^2 =\bm{1}_4, \label{eq:def_I} \\ 
&(C_n)^n = -\bm{1}_4, \label{eq:def_Cn} \\
&S_n^{k_z} = e^{-i \frac{k_z}{2}} C_n, \label{eq:def_Sn} \\
&C_n I = I C_n, \label{eq:rel_CnI} \\ 
&S_n^{-k_z} I = I S_n^{k_z}, \label{eq:rel_SnI}
\end{align}
\end{subequations}
where the minus sign in Eq.~(\ref{eq:def_Cn}) arises due to the $2\pi$ spin rotation.  The screw operation $S_n^{k_z}$ is defined in Eq.~(\ref{eq:def_Sn}) by the combination of the rotation operator $C_n$ and a half translation in the $z$ direction, where the length of the unit cell in the $z$ direction is set to be unity. Since $S^{-\pi}_n=-S^\pi_n$ due to the half-translation factor in Eq.~(\ref{eq:def_Sn}), Eq.~(\ref{eq:rel_SnI}) implies the anti-commutation relation at the Brillouin zone boundary $k_z=\pi$:
\begin{align}
 S_n^{\pi} I = -I S_n^{\pi}. \label{eq:rel_SnI_pi}
\end{align} 
Note that all crystal symmetry operators commute with the TR operator $T$.

In the remainder of this section, we briefly review the classification of 3D Dirac semimetals developed by Yang and Nagaosa~\cite{Yang2014classification}. The symmetry conditions~(\ref{eq:dsm_sym}) and (\ref{eq:dsm_sympro}) allow stable Dirac points to exist on the rotation (screw) axis where two energy bands have different rotation (screw) eigenvalues. Reference~\cite{Yang2014classification} clarified that there are two distinct mechanisms for realizing a stable Dirac point: (a) in rotation symmetric systems, a pair of Dirac points can be formed through an ABC on the rotation axis, which move along the rotation axis and eventually pair annihilate as material parameters are changed. (b) In screw symmetric systems, a \textit{single Dirac point} is pinned at $k_z = \pm \pi$ (i.e., at the Brillouin zone boundaries), which is enforced to exist for any material parameter \cite{YangMorimoto}. 
The difference is attributed to the different algebraic relations with the inversion operation; see Eqs.~(\ref{eq:rel_CnI}) and~(\ref{eq:rel_SnI_pi}). In the following, we will show that the difference between the rotation and screw operations also plays an important role in the classification of GVL phases in $s$-wave superconducting 3D Dirac semimetals.

\subsection{Symmetry in superconducting Dirac semimetals with a vortex line}
\label{sec:sym_SC}
 We consider a superconducting Dirac semimetal in which $s$-wave superconducting order is either intrinsically developed under doping or induced by the proximity effect.
The quasiparticles in the superconducting Dirac semimetal are described by the Bogoliubov-de Gennes (BdG) Hamiltonian,
\begin{equation}
\hat{H}_{\rm BdG} = \frac{1}{2} \sum_{\bm{k}} \sum_{s,s',\sigma,\sigma'}  \Psi^{\dagger}_{\bm{k}, s,\sigma} \tilde{H}_{s,\sigma;s',\sigma'}(\bm{k}) \Psi_{\bm{k}, s',\sigma'},
\end{equation}
where
\begin{equation}
\Psi_{\bm{k},s,\sigma} = \begin{pmatrix}c_{\bm{k},s,\sigma} \\ c_{-\bm{k},s,\sigma}^{\dagger}\end{pmatrix}
\end{equation}
and 
\begin{align}
\widetilde{H}(\bm{k}) = \begin{pmatrix} H(\bm{k})-\mu\bm{1}_4  & \Delta \\ \Delta^{\dagger} & -H^T(-\bm{k})+\mu\bm{1}_4 \end{pmatrix}. \label{eq:BdG}
\end{align}
Here $H(\bm{k})$ is the normal-state Hamiltonian of the Dirac semimetal in Eq.~(\ref{normal Hamiltonian}), $\mu$ is the chemical potential, the superscript $T$ means the transposition, and $\Delta=\Delta_0 (-i s_y \sigma_0) $ describes the $s$-wave pair potential with the superconducting gap $\Delta_0$.  The $s$-wave superconducting Dirac semimetal is a topologically trivial fully-gapped state.  

The symmetry operators on the BdG Hamiltonian are defined by extending Eqs.~(\ref{eq:dsm_sym}) to the Nambu space; the extended TR, inversion, rotation, and screw operators are given by $\widetilde{T} = \diag(T,T^{\ast})$, $\tilde{I} = \diag(I,I^{\ast})$, $\widetilde{C}_n = \diag(C_n,C_n^{ \ast})$, and $\widetilde{S}_n^{k_z} = \diag\biglb(S_n^{k_z},(S_n^{-k_z})^*\bigrb)$, where $*$ is complex conjugation. These operators satisfy the same relations as Eq.~(\ref{eq:dsm_sympro}). In addition, the BdG Hamiltonian is invariant under the particle-hole (PH) transformation:
\begin{align}
C\widetilde{H}(\bm{k})C^{-1} = -\widetilde{H}(-\bm{k}), \quad C=\tau_x \sigma_0 s_0 K \label{eq:phs}
\end{align}
where $\tau_i$ $(i=x,y,z)$ are the Pauli matrices in the Nambu space. The PH operator $C$ commutes with the other operators in the $s$-wave paring state. 
 
Suppose that a vortex line is inserted by external magnetic field applied along the rotation (screw) axis in the superconducting Dirac semimetal. In this situation the spatially varying pair potential is written as $\Delta(\bm{r}) = \Delta(\rho) e^{i\theta}(-is_y\sigma_0)$ in the cylindrical coordinate defined by $\rho \equiv \sqrt{x^2 +y^2}$ and $\theta \equiv \arctan(y/x)$. Here, $\Delta(\rho)$ satisfies $\Delta(0)=0$ and $\Delta(\rho \to \infty)=\Delta_0$. Inserting a vortex line breaks the translation symmetry in the $x$-$y$ plane and the TR symmetry. However, the other crystal symmetries are preserved under the modification explained below.

The BdG Hamiltonian with a vortex line is denoted by $\widetilde{H}_{\rm v}(\rho,\theta,k_z)$.
The $n$-fold rotation operator [Eq.~(\ref{eq:rs})] acts on the coordinates as $(\rho,\theta,k_z) \to (\rho,\theta + \frac{2\pi}{n},k_z)$, which gives an additional phase factor to the pair potential, $\Delta(\rho) e^{i \theta} \to \Delta(\rho) e^{i (\theta + \frac{2\pi}{n})} $. To keep $\widetilde{H}_{\rm v}(\rho,\theta,k_z)$ invariant, we include this phase factor in the $n$-fold rotation operator~\cite{Qin2019}, 
\begin{equation}
\widetilde{C}_{{\rm v}, n}= \begin{pmatrix} e^{i \pi/n} C_n& 0 \\ 0 & e^{-i \pi/n} C_n^{\ast} \end{pmatrix},  \label{eq:rs-vortex}
\end{equation}
such that
\begin{equation}
\widetilde{C}_{{\rm v}, n} \widetilde{H}_{\rm v}(\rho, \theta, k_z) \widetilde{C}_{{\rm v}, n}^{ -1}
= \widetilde{H}_{\rm v}\left(\rho, \theta+ \frac{2\pi}{n}, k_z\right).
\end{equation}
In a similar manner, the $n$-fold screw operator for the BdG Hamiltonian $\widetilde{H}_{\rm v}(\rho,\theta,k_z)$ is given by
\begin{equation}
\widetilde{S}_{{\rm v},n}^{k_z} = \begin{pmatrix}e^{i\pi/n}S^{k_z}_n & 0 \\ 0 & e^{-i\pi/n}(S^{-k_z}_n)^* \end{pmatrix}.
\end{equation}
In addition, since the inversion operation acts on the coordinates as $ (\rho,\theta,k_z) \to  (\rho,\theta+\pi,-k_z)$, we define the inversion operation in the presence of a vortex line as
\begin{equation}
\tilde{I}_{\rm v} = \begin{pmatrix} e^{i \pi/2} I & 0 \\ 0 & e^{-i \pi/2} I^{\ast} \end{pmatrix} , \label{eq:is-vortex} 
\end{equation}
such that
\begin{equation}
\tilde{I}_{\rm v} \widetilde{H}_{\rm v}(\rho, \theta, k_z) \tilde{I}_{\rm v}^{ -1} = \widetilde{H}_{\rm v}\left(\rho, \theta+ \pi, -k_z\right).
\end{equation}
The symmetry properties of the modified operators are summarized as follows:
\begin{subequations} \label{eq:gvl_sympro}
\begin{align}
 &(\tilde{I}_{\rm v})^2 = -\bm{1}_8, \label{eq:rel_I_vortex}  \\
 &(\widetilde{C}_{{\rm v}, n})^n = \bm{1}_8, \label{eq:rel_Cn_vortex}  \\
 &\widetilde{C}_{{\rm v}, n}\tilde{I}_{\rm v} = \tilde{I}_{\rm v} \widetilde{C}_{{\rm v}, n}, \label{eq:rel_CnI_vortex} \\
 &\widetilde{S}_{{\rm v}, n}^{-k_z} \tilde{I}_{\rm v} = \tilde{I}_{\rm v} \widetilde{S}_{{\rm v}, n}^{k_z}. \label{eq:rel_SnI_vortex} 
\end{align}
\end{subequations}
We notice that the right-hand side of Eqs.~(\ref{eq:rel_I_vortex}) and (\ref{eq:rel_Cn_vortex}) has the opposite sign to Eqs.~(\ref{eq:def_I}) and (\ref{eq:def_Cn}), respectively, due to the extra phase factors.
Equation (\ref{eq:rel_SnI_vortex}) implies
\begin{equation}
\widetilde{S}^\pi_{{\rm v},n}\tilde{I}_{\rm v}=-\tilde{I}_{\rm v}\widetilde{S}^\pi_{{\rm v},n}.
\label{tildeStildeI=-tildeItildeS}
\end{equation}
On the other hand, the PH symmetry operator, already defined in Eq.~(\ref{eq:phs}), satisfies
\begin{align}
 C\widetilde{H}_{\rm v}(\rho,\theta,k_z)C^{-1} =  -\widetilde{H}_{\rm v}(\rho,\theta,-k_z),
 \end{align}
 and commutes with the modified operators.

\subsection{Effective low-energy Hamiltonian}
\label{sec:basis_principle}

Here, we explain our strategy for classifying possible GVL phases. Key ingredients are the symmetry relations~(\ref{eq:gvl_sympro}) in the presence of a vortex line, which determine the stability of GVL phases.
Since the gapless modes in a vortex are extended along the $k_z$ direction and localized in the radial ($\rho$) direction,
the topology of GVL phases can be understood from a 1D BdG Hamiltonian~\cite{Kobayash2020}, $\mathcal{H}(k_z)$, which is obtained from $\widetilde{H}_{\rm v}(\rho, \theta, k_z)$ by regarding $(\rho,\theta)$ as internal indices and taking only $k_z$ as the relevant spatial direction.
Thus we can classify GVL modes as 1D nodal superconducting phases.

The symmetry operators of $\mathcal{H}(k_z)$ are the PH ($\mathcal{C}$), inversion ($\widetilde{\mathcal{I}}$), and $n$-fold rotation or screw ($\widetilde{\mathcal{C}}_n$ or $\widetilde{\mathcal{S}}_n^{k_z}$) symmetries, which satisfy Eq.~(\ref{eq:gvl_sympro}).
The 1D Hamiltonian for GVL modes $\mathcal{H}(k_z)$ commutes with $\widetilde{\mathcal{C}}_n$ ($\widetilde{\mathcal{S}}_n^{k_z}$), and the energy levels are labeled by the rotation (screw) eigenvalues. In addition, the relation (\ref{tildeStildeI=-tildeItildeS}) leads to two-fold degeneracy at $k_z=\pi$, which is important for the classification of GVL phases under the screw symmetry.

We start with symmetry analysis at the high-symmetry points $k_{\rm inv}=0$ and $\pi$. Since $\mathcal{H}(k_{\rm inv})$ commutes with the crystal symmetry operators, $\mathcal{H}(k_{\rm inv})$ can be decomposed into block matrices in terms of irreducible representations (irreps) of the symmetry operators. Vortex bound states form a basis of the irreps, and the dimension of irreps gives constraints on the number of energy bands that we need to describe the GVL modes. Furthermore, the PH symmetry demands that irreps for electron and hole states come together in any matrix of the symmetry operators.  Hence, $\mathcal{H}(k_{\rm inv})$ and symmetry operators are decomposed as
\begin{align}
 &\mathcal{H}(k_{\rm inv}) \simeq \oplus_{\alpha} \mathcal{H}_\alpha(k_{\rm inv}), \\
 &\mathcal{C} \simeq \oplus_{\alpha} \begin{pmatrix} 0 & \bm{1}_{n_{\alpha}} \\ \bm{1}_{n_{\alpha}} & 0 \end{pmatrix} K , \\
&\widetilde{\mathcal{C}}_{n} \simeq  \oplus_{\alpha} \begin{pmatrix}\mathcal{C}_{n,\alpha} & 0 \\ 0 & \mathcal{C}_{n,\alpha}^{\ast}  \end{pmatrix}, \\
  &\widetilde{\mathcal{S}}_{n}^{k_{\rm inv}} \simeq \oplus_{\alpha} \begin{pmatrix}\mathcal{S}_{n,\alpha}^{k_{\rm inv}} & 0 \\ 0 & \mathcal{S}_{n,\alpha}^{-k_{\rm inv} \ast}  \end{pmatrix}, \\
  &\tilde{\mathcal{I}} \simeq \oplus_{\alpha} \begin{pmatrix} \mathcal{I}_{\alpha} & 0 \\ 0 & \mathcal{I}_{\alpha}^{\ast} \end{pmatrix}, 
\end{align}
where $\alpha$ is a label of irreps, $n_{\alpha}$ is the dimension of irrep $\alpha$, and $\simeq$ means the equivalence under a unitary transformation. 
We pick up one block matrix as a minimal Hamiltonian describing vortex bound states at low energies ($|E| \ll \Delta_0$), which can be described as
\begin{align}
 \mathcal{H}_\alpha (k_{\rm inv}) = \sum_i a_i (k_{\rm inv}) M_i, \label{eq:lowHami}
\end{align}
where $M_i$ $(i=1, \dots, 4n^2_{\alpha} )$ are $2n_{\alpha} \times 2n_{\alpha} $ matrices given by the direct product of Pauli matrices.
To find the dispersion relation around the high symmetry points, we expand $a_i(k_z)$ at $k_z=k_{\rm inv}$ to the leading order of $k_z$ under the symmetry constraints.

\subsection{Classification of vortex-line phases}
\label{sec:classification}

\begin{figure*}[tbp]
\centering
 \includegraphics[width=16cm]{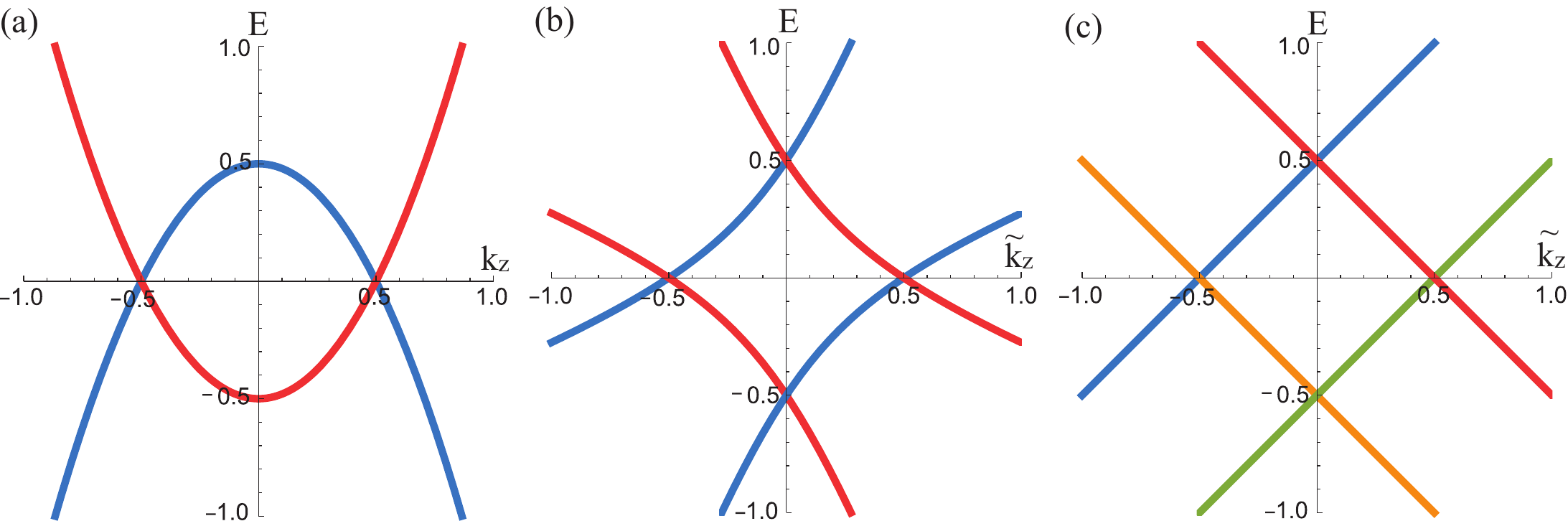}
 \caption{ (Color online) Energy spectrum of (a) Eq.~(\ref{eq:rot_Ev2}), (b) Eq.~(\ref{eq:screw_E2}), and (c) Eq.~(\ref{eq:screw_E3}), for the parameters $(m_0,m_1)=(0.5,-2)$, $(m_1,v_1,v_2,v_3)=(0.5,1.5,1,0.5)$, and $(m_0,v_1)=(0.5,1)$, respectively. The colors of the lines represent the rotation (screw) eigenvalues. The red and blue lines indicate $\exp(i\frac{2\pi p}{n})$ and $\exp(-i\frac{2\pi p}{n})$ in (a), and $-1$ and $1$ in (b); the red, blue, orange, and green lines, respectively, represent $-\exp[i(\frac{2\pi p}{n}-\frac{\pi}{2})]$, $\exp[i(\frac{2\pi p}{n}-\frac{\pi}{2})]$, $-\exp[-i(\frac{2\pi p}{n}-\frac{\pi}{2})]$, and $\exp[-i(\frac{2\pi p}{n}-\frac{\pi}{2})]$ in (c). In (b) and (c), the bands are two-fold degenerate at $k_z=\pi$ $(\tilde{k}_z=0$) due to the screw and inversion symmetries.}
\label{fig:kp}
\end{figure*}

We now construct the symmetry-adopted expression of $\mathcal{H}_\alpha (k_z)$ under the symmetry constraints and determine possible gapless energy modes.  
Since the rotation symmetry is independent of $k_z$ and $\mathcal{S}_{n,\alpha}^{0} = \mathcal{C}_{n,\alpha}$, we only need to consider the following two cases: vortex-line modes (a) around $k_{\rm inv} = 0$ protected by $\mathcal{C}_{n,\alpha}$ and (b) around $k_{\rm inv} = \pi$ protected by $\mathcal{S}_{n,\alpha}^{\pi}$. In the following, we address possible GVL modes for the cases (a) and (b).

\subsubsection{Rotation-symmetry-protected GVL modes}
For the case (a), we consider the rotation operator at $k_z=0$ that commutes with the inversion operator. The irrep is one dimensional ($n_{\alpha}=1$). The crystal symmetry operators satisfy Eqs.~(\ref{eq:rel_I_vortex}) and~(\ref{eq:rel_Cn_vortex}), so that the the irreps can be written as
\begin{align}
 \mathcal{C}_{n,p} = e^{i \frac{2\pi p}{n}}, \quad \mathcal{I}_{\rm R}=i, \label{eq:rot_irrep}
 \end{align}
 and the representations in the Nambu space are
\begin{equation}
 \widetilde{\mathcal{C}}_{n,p} =  e^{i \frac{2\pi p}{n} \tau_z}, \quad \widetilde{\mathcal{I}}_{\rm R} =i \tau_z, \label{eq:op_comm}
\end{equation}
where $p=1,\cdots, n$ label irreps of the rotation operator. They satisfy $[\mathcal{C},\widetilde{\mathcal{C}}_{n,p} ]=[\mathcal{C},\widetilde{\mathcal{I}}_{\rm R}]=0$ with the PH operator $\mathcal{C}=\tau_x K$.
Since the Hamiltonian is a $2 \times 2$ matrix, it can be expanded by the Pauli matrices:
\begin{align}
\mathcal{H}_{n,p}^{\rm R}(k_z) &= \sum_{i=x,y,z} a_{i}(k_z) \tau_i,  \ \ a_i \in \mathbb{R}, \label{eq:Heff_rot}
\end{align}
where $\tau_0=\bm{1}_2$ is forbidden due to the $\mathcal{C}\tilde{\mathcal{I}}_{\rm R}$ symmetry. 
The inversion symmetry, $\widetilde{\mathcal{I}}_{\rm R}\mathcal{H}_{n,p}^{\rm R}(k_z) \widetilde{\mathcal{I}}^{-1}_{\rm R} = \mathcal{H}_{n,p}^{\rm R}(-k_z)$, constrains the coefficients $a_{x}$ and $a_{y}$ to be odd functions of $k_z$, and $a_{z}$ to be an even function of $k_z$. The rotation symmetry, $[\widetilde{\mathcal{C}}_{n,p},\mathcal{H}_{n,p}^{\rm R}(k_z) ] = 0$, imposes an extra condition
\begin{align}
 &e^{\mp i \frac{4 \pi p}{n}} a_{\pm} (k_z) = a_{\pm}(k_z),
\end{align}  
where we have used $\widetilde{\mathcal{C}}_{n,p}\tau_{\pm}  \widetilde{\mathcal{C}}_{n,p}^{-1} = e^{\pm i 4 \pi p/n}\tau_{\pm} $ with $\tau_{\pm} = \tau_x \pm i \tau_y$ and $a_{\pm}  = (a_x \pm i a_y)/2$.
We have the following two possibilities depending on whether $2p/n \in \mathbb{Z}$ or not. 

(i) $2p/n \in \mathbb{Z}$:  
$a_{\pm}(k_z)$ is generally nonzero, and the energy eigenvalues are
\begin{align}
E_{\pm}^{\rm R1}(k_z) = \pm \sqrt{ a_{x}^2(k_z) + a_{y}^2(k_z) + a_{z}^2(k_z)} \, . \label{eq:rot_Ev1}
\end{align}
This energy spectrum is fully gapped in general, since the condition $a_x=a_y=a_z=0$ cannot be satisfied simultaneously when we only have one parameter $k_z$.

(ii) $2p/n \notin \mathbb{Z}$: the rotation symmetry imposes $a_{\pm}(k_z)=0$, and the energy spectrum in the leading order of $k_z$ is then given by
\begin{align}
 E_{\pm}^{\rm R2}(k_z) = \pm a_z(k_z) = \pm (m_0+m_1 k_z^2),  \label{eq:rot_Ev2}
\end{align}
where $m_0,m_1 \in \mathbb{R}$. Thus, a pair of gapless points can appear due to ABC at 
\begin{align}
k_z = \pm \sqrt{-m_0/m_1}, \label{eq:rot_kz}
\end{align}
when $m_0/m_1 <0$; see Fig.~\ref{fig:kp} (a).
We conclude that the GVL phases are possible when the rotation eigenvalues are complex values,  in agreement with the previous works~\cite{Qin2019,Konig2019}.

 \begin{table}[tb]
\caption{
Classification of GVL phases for $2 \times 2$ Hamiltonians with the rotation symmetry around  $k_z=0$. The Hamiltonians with the rotation symmetry around $k_z=\pi$ and those with the screw symmetry around $k_z=0$ also have the same classification.  The first and second columns indicate the $n$-fold rotation operators, where $n$ and $p$ are defined by Eq.~(\ref{eq:rot_irrep}). The third, forth, and fifth columns show the types of the Hamiltonians, the vortex-line phases, and the underlying mechanism for the formation of gapless modes. Here, ABC stands for accidental band crossing.  
}
\label{tab:rot_class}
\begin{tabular}{ccccc}
\hline\hline
$n$ & $p$ & Type &Phase & Mechanism\\
\hline 
$2$ & 1,2  & (i)  &Gapped  &\\
$3$ & 3   &  (i)   &Gapped &\\
$3$ & 1,2 &  (ii)& Gapless or Gapped & ABC \\
$4$ & 2,4 &  (i)  &Gapped &\\
$4$ & 1,3 &  (ii) & Gapless or Gapped & ABC \\
$6$ & 3,6 &  (i)   &Gapped& \\
$6$ & 1,2,4,5 &  (ii) & Gapless or Gapped & ABC \\
\hline\hline
\end{tabular} 
\end{table} 

\begin{table}[tb]
\caption{
Classification of GVL phases for $4 \times 4$ Hamiltonians with the screw symmetry around  $k_z=\pi$. The first and second columns indicate the $n$-fold screw operators defined by Eq.~(\ref{eq:screw_irrep}). 
The third, forth, and fifth columns are the same as Table~\ref{tab:rot_class}. 
Here, EBC stands for enforced band crossing.  
}
\label{tab:screw_class}
\begin{tabular}{ccccc}
\hline\hline
$n$ & $p$ & Type &Phase & Mechanism\\
\hline 
$2$ & 1  & (iii) & Gapped& \\
$4$ & 2 &  (iii) & Gapped & \\
$4$ &1,3 &  (iv) & Gapless or Gapped & ABC \\
$6$ & 3 &  (iii) & Gapped & \\
$6$ & 1,2,4,5 &  (v) &Gapless & EBC \\
\hline\hline
\end{tabular} 
\end{table}

\subsubsection{Screw-symmetry-protected GVL modes}
For the case (b), we consider the screw operator at $k_z=\pi$ that anti-commutes with the inversion operator. The irrep is two dimensional ($n_{\alpha} =2$). On the basis that diagonalize the screw operator, the screw and inversion operators are represented by 
\begin{equation}
\mathcal{S}_{n,p}^{\pi} =  -i e^{i \frac{2\pi p}{n}} \sigma_z, \quad \mathcal{I}_{\rm S}= i \sigma_x, \label{eq:screw_irrep}
\end{equation}
where $n=2,4,6$ and $p=1,\cdots,n/2$. $\sigma_i$ ($i=x,y,z$) are the Pauli matrices stemming from the sublattice degrees of freedom associated with the screw operation~\footnote{Note that we can choose other representations, say $\mathcal{S}_{n,p}^{\pi} =  -i \exp(i \frac{2\pi p}{n}) \sigma_z$ and $\mathcal{I} = i\sigma_y$, which are related to Eq.~\ref{eq:screw_irrep} by a unitary transformation. Thus, we obtain the same energy eigenvalues.}. The factor $-i$ in the screw operator comes from a half translation. The representations in the Nambu space are then given by
\begin{equation}
 \widetilde{\mathcal{S}}_{n,p}^{\pi} = e^{i \left( \frac{2\pi p}{n} -\frac{\pi}{2}\right) \tau_z} \sigma_z , \quad
 \widetilde{\mathcal{I}}_{\rm S}= i \tau_z \sigma_x, \label{eq:op_anticomm}
\end{equation}
which satisfy $[\mathcal{C},\widetilde{\mathcal{C}}_{n,p} ]=[\mathcal{C},\widetilde{\mathcal{I}}_{\rm S}]=0$ with the PH operator $\mathcal{C}=\tau_x \sigma_0 K$. 
The effective Hamiltonian can be expanded by $4 \times 4$ matrices consisting of two Pauli matrices $\sigma_i$ and $\tau_i$, 
\begin{align}
\mathcal{H}_{n,p}^{\rm S}(\tilde{k}_z) = &\sum_{i,j=0,x,y,z} a_{ij}(\tilde{k}_z) \tau_i \sigma_j, \ \ a_{ij} \in \mathbb{R},
\end{align}
where $\tilde{k}_z \equiv k_z -\pi$.
The (anti-)symmetry of the Hamiltonian under the $\mathcal{C}\widetilde{\mathcal{I}}_{\rm S}$ operation, i.e., $\{\mathcal{C}\widetilde{\mathcal{I}}_{\rm S},\mathcal{H}_{n,p}^{\rm S}(\tilde{k}_z) \}=0$, imposed constraints on the Hamiltonian: $a_{00}=a_{0x}=a_{0y}=a_{xz}=a_{yz}=a_{zz}=0$. The remaining ten coefficients are further constrained by the crystalline symmetries. First, the inversion symmetry,  $\widetilde{\mathcal{I}}_{\rm S}\mathcal{H}_{n,p}^{\rm S}(\tilde{k}_z) \widetilde{\mathcal{I}}^{-1}_{\rm S} = \mathcal{H}_{n,p}^{\rm S}(-\tilde{k}_z)$, determines the momentum dependence of $a_{ij}$:  the four coefficients ($a_{z0}$, $a_{zx}$, $a_{xy}$, $a_{yy}$) are even functions of $k_z$ and the others ($a_{0z}$, $a_{zy}$, $a_{x0}$, $a_{xx}$, $a_{y0}$, $a_{yx}$) are odd functions of $k_z$.
Second, the screw symmetry, $[\tilde{\mathcal{S}}_{n,p}^{\pi} ,\mathcal{H}_{n,p}^{\rm S}(k_z)]=0$, gives the following constraints on the coefficients:
\begin{subequations}\label{eq:screw_rel}
\begin{align}
 &a_{zx}(\tilde{k}_z) = 0, \label{eq:screw_rel3} \\
 &a_{zy}(\tilde{k}_z) = 0, \label{eq:screw_rel4} \\
 &-e^{\mp i\frac{4\pi p}{n}} a_{\pm 0}(\tilde{k}_z) = a_{\pm 0}(\tilde{k}_z), \label{eq:screw_rel5} \\
 &e^{\mp i\frac{4\pi p}{n}} a_{\pm x}(\tilde{k}_z) = a_{\pm x}(\tilde{k}_z), \label{eq:screw_rel6} \\
 &e^{\mp i\frac{4\pi p}{n}} a_{\pm y}(\tilde{k}_z) = a_{\pm y}(\tilde{k}_z), \label{eq:screw_rel7} 
\end{align}
\end{subequations}
where $a_{\pm i} \equiv (a_{xi} \pm i a_{yi})/2$ and we have used the relations
\begin{subequations}
\begin{align}
&\mathcal{S}_{n,p}^{\pi} \tau_{\pm} \sigma_0 (\mathcal{S}_{n,p}^{\pi})^{-1} =- e^{\pm i\frac{4\pi p}{n}}\tau_{\pm} \sigma_0, \\
&\mathcal{S}_{n,p}^{\pi} \tau_{\pm} \sigma_{x} (\mathcal{S}_{n,p}^{\pi})^{-1} = e^{\pm i\frac{4\pi p}{n}}\tau_{\pm} \sigma_{x}, \\
&\mathcal{S}_{n,p}^{\pi} \tau_{\pm} \sigma_{y} (\mathcal{S}_{n,p}^{\pi})^{-1} = e^{\pm i\frac{4\pi p}{n}}\tau_{\pm} \sigma_{y}.
\end{align}
\end{subequations}
Since Eqs.~(\ref{eq:screw_rel5}),  (\ref{eq:screw_rel6}), and~(\ref{eq:screw_rel7}) depend on $n$ and $p$, the possible forms of the Hamiltonian are discussed separately for three cases as below. 

(iii) $\widetilde{\mathcal{S}}_{n,p}^{\pi} = \pm i \tau_z  \sigma_z$ when $2p/n \in \mathbb{Z}$, i.e., $(n,p)=(2,1)$, (4,2), or (6,3); $a_{\pm0}=0$ from Eq.~(\ref{eq:screw_rel5}). Thus, the Hamiltonian is given by 
\begin{align}
 \mathcal{H}_{n,p}^{\rm S}(\tilde{k}_z) =&\, a_{0z} \tau_0 \sigma_z +a_{z0} \tau_z \sigma_0 \nonumber \\
           &+( a_{+ x} \tau_{-} \sigma_x + a_{+ y} \tau_{-} \sigma_y + {\rm H.c.}),
\end{align}
whose energy eigenvalues are 
\begin{align}
& E_{s,r}^{\rm S1}(\tilde{k}_z) \nonumber \\
 &= s \sqrt{(a_{0z}+ r a_{z0})^2+(a_{xx} - r a_{yy})^2+(a_{xy} + r a_{yx})^2}, \label{eq:screw_E1}
\end{align}
where $s,r \in \{\pm 1\}$. This leads to fully-gapped energy spectra since the gapless condition $E_{s,r}^{\rm S1}(\tilde{k}_z) =0$ requires $a_{0z}=-r a_{z0}$, $a_{xx} = r a_{yy}$, and $a_{xy} = -r a_{yx}$, which cannot be satisfied simultaneously by tuning the single parameter $k_z$.

 (iv) $\widetilde{\mathcal{S}}_{n,p}^{\pi} = \pm \tau_0  \sigma_z$ when $(4p- n)/2n \in \mathbb{Z}$, i.e., $(n,p)=(4,1)$ or (4,3); we find $a_{\pm x}=a_{\pm y}=0$ due to the constraints~(\ref{eq:screw_rel6}) and (\ref{eq:screw_rel7}).
The effective Hamiltonian becomes 
\begin{align}
 \mathcal{H}_{n,p}^{\rm S}(\tilde{k}_z)  = a_{0z} \tau_0 \sigma_z +a_{z0} \tau_z \sigma_0  
 + (a_{+0} \tau_{-} \sigma_0 +{\rm H.c.}) \label{eq:screw_H2}
\end{align}
with the eigenvalues
\begin{align}
 E_{s,r}^{\rm S2}(\tilde{k}_z)= &s a_{0z}+ r \sqrt{a_{x0}^2+ a_{y0}^2+a_{z0}^2}. \label{eq:screw_E2}
\end{align}
In this case, we can find a solution satisfying $E_{s, r}^{\rm S2}(\tilde{k}_z)=0$ as follows.
We expand the coefficients in lowest order in $\tilde{k}_z$: $a_{0z}=v_1\tilde{k}_z$, $a_{z0}=m_0$, $a_{x0}=v_2\tilde{k}_z$, and $a_{y0} = v_3\tilde{k}_z$ ($m_0, v_1,v_2,v_3 \in \mathbb{R}$).
The solutions of $E_{\pm, \mp}^{\rm S2}(\tilde{k}_z)=0$ or $E_{\pm, \pm}^{\rm S2}(\tilde{k}_z)=0$ are obtained as
\begin{align}
\tilde{k}_z  = \pm \frac{m_0}{\sqrt{v_1^2-v_2^2-v_3^2}}, \label{eq:screw_kz2}
\end{align}
which means that the ABC occurs in the parameter region satisfying $v_1^2-v_2^2-v_3^2 > 0$ as shown in Fig.~\ref{fig:kp} (b).

 (v) the other cases:  Eqs.~(\ref{eq:screw_rel5}),  (\ref{eq:screw_rel6}), and~(\ref{eq:screw_rel7}) give $a_{\pm 0}=a_{\pm x}=a_{\pm y}=0$. Thus, the Hamiltonian consists of two terms, 
\begin{align}
 \mathcal{H}_{n,p}^{\rm S}(\tilde{k}_z)  = a_{0z} \tau_0  \sigma_z +a_{z0} \tau_z \sigma_0,
\end{align}
with the energy eigenvalues
\begin{align}
E_{s,r}^{\rm S3}(\tilde{k}_z)= s a_{0z} + r a_{z0}. \label{eq:screw_E3}
\end{align}
The solution of $E_{s,r}^{\rm S3}(\tilde{k}_z)=0$ is given by
\begin{equation}
 \tilde{k}_z = \pm \frac{m_0}{v_1}, \label{eq:screw_kz3}
\end{equation}
where we put $a_{0z}=v_1\tilde{k}_z$ and $a_{z0}=m_0$ ($m_0,v_1\in \mathbb{R}$). The energy spectra are shown in See Fig.~\ref{fig:kp} (c).  Interestingly, the gapless mode appears whenever $v_1 \neq 0$, more robustly than the other cases. We call this case an EBC. 

In summary, the GVL phases are realized for both rotation and screw symmetries. Their classification is summarized in Table~\ref{tab:rot_class} and~\ref{tab:screw_class}, in which we find that the rotation and screw symmetry-protected GVL phases can appear for the same sets of $n$ and $p$. This correspondence and the relation $\mathcal{C}_{n,p}=\mathcal{S}^0_{n,p}$ imply that the gapless points can be moved from $k_z =\pi$ to the vicinity of $k_z=0$ along the screw symmetric line by tuning model parameters. 
The similar results are obtained in the K-theoretical classification, which is discussed in Appendix~\ref{app:k-th}. 

\section{Vortex gapless modes protected by screw symmetry}
\label{sec:screw}
\subsection{Tight-binding model}
\label{sec:model}

\begin{figure}[tp]
\centering
 \includegraphics[width=8.5cm]{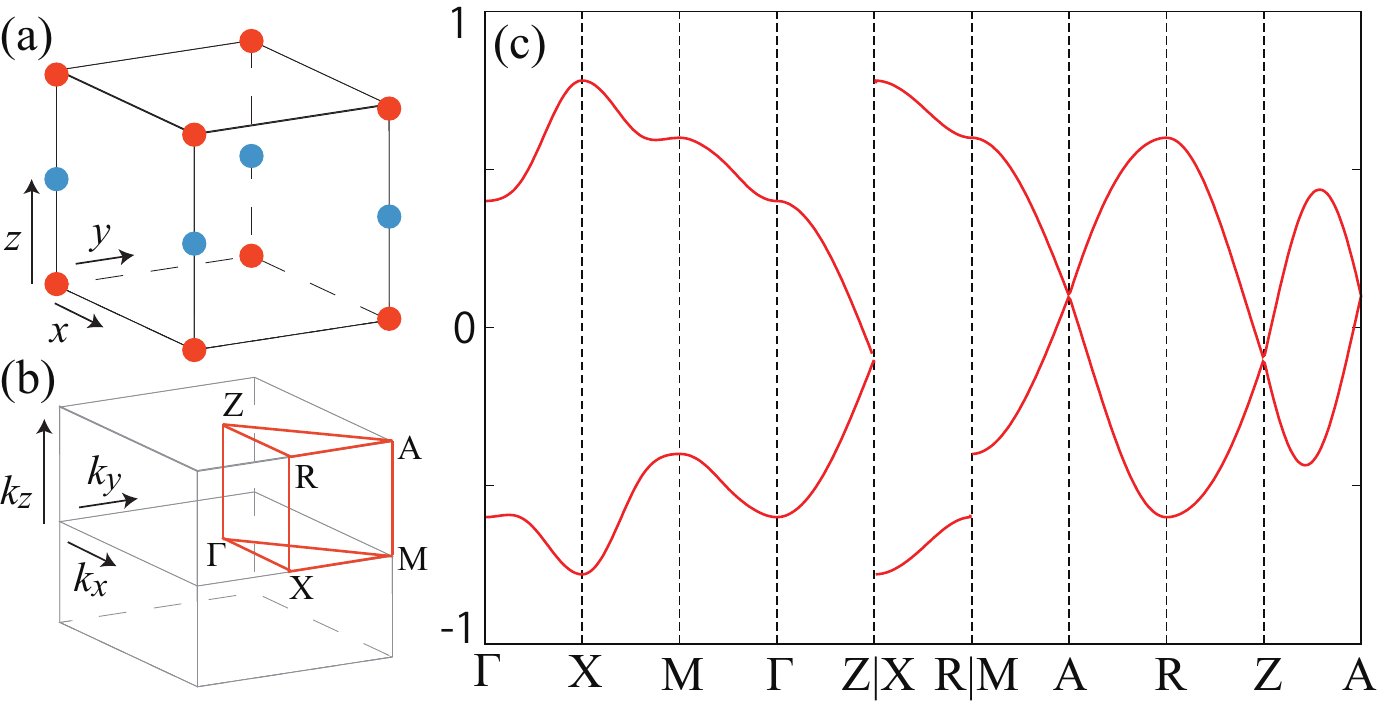}
 \caption{ (Color online) (a) Tetragonal lattice structure with two sites in a unit cell. (b) Brillouin zone of the tetragonal lattice, where the red thick lines represent high symmetry lines. (c) Energy spectrum of the Hamiltonian (\ref{eq:tbmodel}) along the high symmetry lines. Every energy band is doubly degenerate. There are two Dirac points positioned at $Z=(0,0,\pi)$ and $A=(\pi,\pi,\pi)$. The parameters are chosen to be $(t_1,t_{xy},t_z,\lambda_1,\lambda_2)=(-0.05,0.3,0.5,0.3,0.1)$. }
\label{fig:band}
\end{figure}

In this section we demonstrate a screw-symmetry-protected GVL phase which, to our knowledge, has not been discussed before. We consider a model on a tetragonal lattice with the space group symmetry $P4_2/mmc$ (SG\#{}131). The lattice structure is shown in Fig.~\ref{fig:band}(a). The unit cell has two sites located at $z=0$ and $z=1/2$ along the $z$ axis, where the length of the unit cell is set to be unity in the $x$, $y$, and $z$ directions.  We assume that an $s$-orbital electron resides on each site. 
Our model realizes the type-(iv) GVL phase discussed in Sec.~\ref{sec:classification}.

The tight-binding Hamiltonian in the normal state is written as
 \begin{align}
  H(\bm{k}) = & \, t_1 [\cos(k_x) + \cos (k_y)]\bm{1}_{4}  \nonumber \\
          &+ t_{xy}  [\cos(k_x)-\cos(k_y)]\sigma_z s_0 \nonumber \\
          &+\frac{t_z}{2} [\sigma_x + \sigma_x(2k_z)] s_0 \nonumber \\
  &+\frac{\lambda_1}{2} [\sigma_x - \sigma_x(2k_z)] [s_x \sin(k_y)+s_y \sin(k_x)] \nonumber \\
  & + \frac{\lambda_2}{2} [\sigma_y + \sigma_y(2k_z)] s_z \sin(k_x) \sin(k_y), \label{eq:tbmodel}
 \end{align}
where $t_1$, $t_{xy}$, and $t_z$ are hopping matrix elements, $\lambda_1$ and $\lambda_2$ are spin-orbit couplings,
$s_i$ and $\sigma_i$ ($i=x,y,z$) are the Pauli matrices in the spin and sublattice spaces, respectively, and
the $k_z$-dependent Pauli matrices are defined by
 \begin{align}
  & \sigma_0(k_z) = \begin{pmatrix}  e^{-i \frac{k_z}{2}} & 0 \\ 0 & e^{i \frac{k_z}{2}} \end{pmatrix}, \\
  & \sigma_x(k_z) = \begin{pmatrix} 0 & e^{i \frac{k_z}{2}} \\ e^{-i \frac{k_z}{2}} & 0\end{pmatrix}, \\
  & \sigma_y(k_z) = \begin{pmatrix} 0 & -ie^{i \frac{k_z}{2}} \\ i e^{-i \frac{k_z}{2}} & 0\end{pmatrix}.
 \end{align}
 
 \begin{figure*}[tbp]
\centering
 \includegraphics[width=17cm]{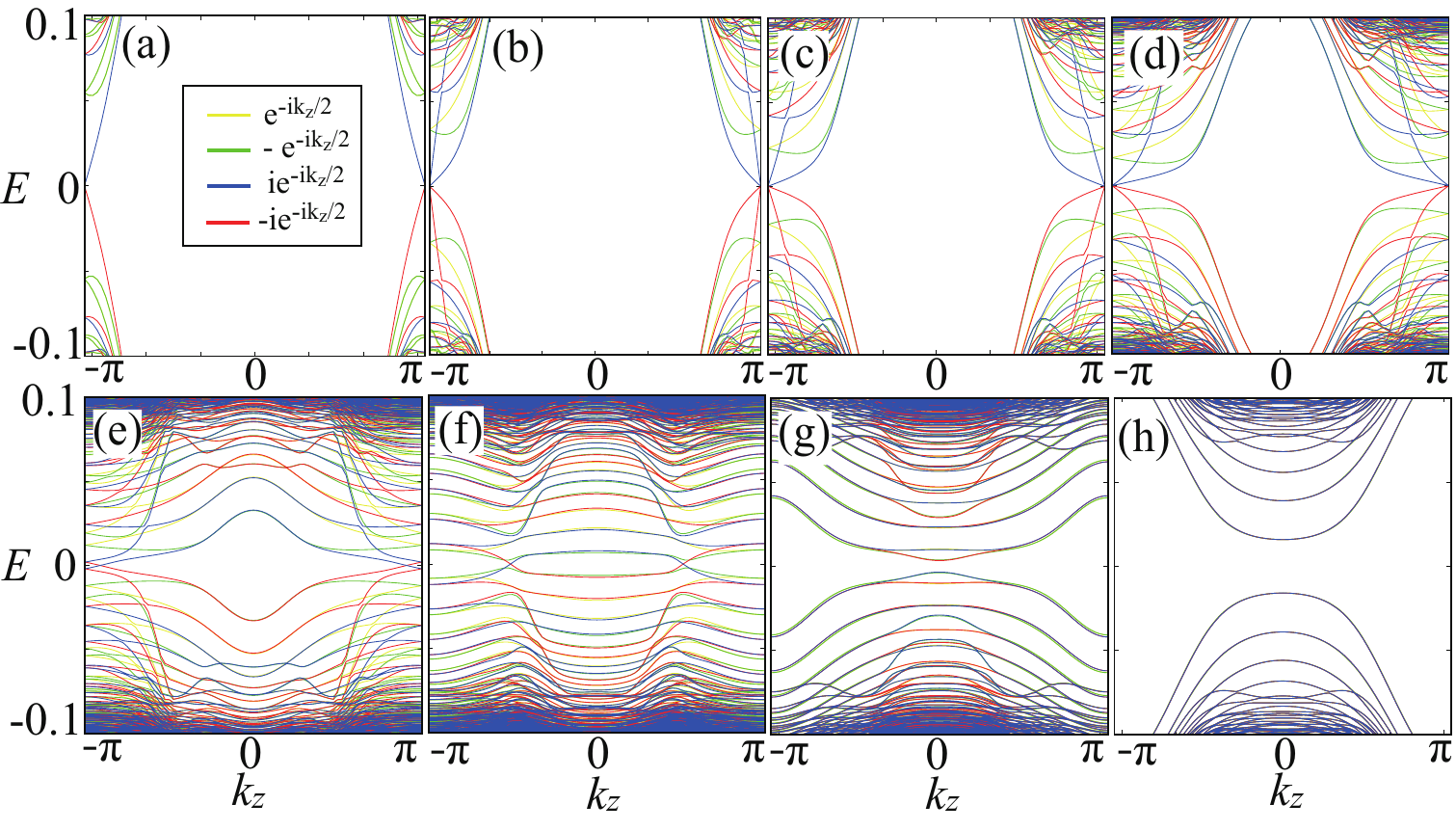}
 \caption{ (Color online) Energy levels as a function of $k_z$ of vortex bound states of the BdG Hamiltonian (\ref{eq:BdG_tbm}), constructed from the normal-state Hamiltonian (\ref{eq:tbmodel}) and the $s$-wave pair potential with a vortex line (\ref{eq:vortexline}): (a) $\mu=0$, (b) $-0.1$, (c) $-0.2$, (d) $-0.3$, (e) $-0.4$, (f) $-0.5$, (g) $-0.6$, and (h) $-0.7$. Here, we choose the same hopping parameters as in Fig.~\ref{fig:band} and the pair potential with $(\Delta_0,\xi)=(0.1,4)$. The system size is $L=21$. The inset of (a) indicates the screw eigenvalues: the yellow, green, blue, and red lines have the screw eigenvalues $-i$, $i$, $1$, and $-1$ at $k_z=\pi$, respectively. Following the argument in Sec.~\ref{sec:classification}, the yellow and green lines (the red and blue lines) can be described by the effective Hamiltonian $\mathcal{H}_{4,2}^{\rm S}(\tilde{k}_z)$ [$\mathcal{H}_{4,1}^{\rm S}(\tilde{k}_z)$]. Thus, the GVL modes are allowed in the red and blue lines.}
\label{fig:tbvm}
\end{figure*}
 
The Hamiltonian preserves the TR symmetry ($T= i \sigma_0 s_y K$) and the crystal symmetry of $P4_2/mmc$, which comprises four-fold screw symmetry about the $z$ axis ($S_{4z}^{k_z}$), two-fold rotation symmetry about the $x$ axis ($C_{2x}^{k_z}$), and inversion symmetry ($I^{k_z}$).
The Hamiltonian is invariant under these crystal symmetry operations,
 \begin{align}
 &S_{4z}^{k_z}H(\bm{k})\bigl(S_{4z}^{k_z}\bigr)^{-1} = H(R_4\bm{k}), \label{eq:s4z_tbm}  \\
 &C_{2x}^{k_z} H(k_x,k_y,k_z) \bigl(C_{2x}^{k_z}\bigr)^{-1} = H(k_x,-k_y,-k_z), \label{eq:c2x_tbm} \\
 &I^{k_z} H(\bm{k}) \bigl(I^{k_z}\bigr)^{-1} = H(-\bm{k}),  \label{eq:I_tbm}
 \end{align}
 where the operators are defined by
 \begin{align}
 &S_{4z}^{k_z} = e^{-i \frac{k_z}{2}} \sigma_x(k_z) e^{-i\frac{\pi}{4} s_z}, \\
 &C_{2x}^{k_z} = i e^{i \frac{k_z}{2}}  \sigma_{0}(k_z)s_x,   \\
 & I^{k_z}= e^{i \frac{k_z}{2}}  \sigma_{0}(k_z) s_0.
 \end{align}
The two-fold rotation and inversion operations satisfy $C_{2x}^{-k_z}C_{2x}^{k_z}=-\bm{1}_4$ and $I^{-k_z}I^{k_z}=\bm{1}_4$.
The commutation relation between $\mathcal{S}_{4z}^{k_z}$ and $I^{k_z}$ is given by 
 \begin{align}
  \mathcal{S}_{4z}^{-k_z}\mathcal{I}^{k_z} = \mathcal{I}^{k_z} \mathcal{S}_{4z}^{k_z},
 \end{align}
which leads to the anti-commutation relation at $k_z=\pi$,
\begin{equation}
\{\mathcal{S}_{4z}^\pi, \mathcal{I}^\pi\} = 0.
\end{equation}
The band structure of Eq.~(\ref{eq:tbmodel}) is shown in Fig.~\ref{fig:band}(c), in which all the energy bands are doubly degenerate due to the presence of TR and inversion symmetries. The two Dirac points at $Z=(0,0,\pi)$ and $A=(\pi,\pi,\pi)$ are protected by the four-fold screw symmetry [Eq.~(\ref{eq:s4z_tbm})] and the inversion symmetry [Eq.~(\ref{eq:I_tbm})].

We extend the normal-state Hamiltonian [Eq.~(\ref{eq:tbmodel})] to the BdG Hamiltonian
 \begin{align}
 \widetilde{H}(\bm{k})  = \begin{pmatrix} H(\bm{k}) -\mu \bm{1}_4 & \Delta \\ \Delta^{\dagger} &  -H^T(-\bm{k}) +\mu \bm{1}_4 \end{pmatrix}, \label{eq:BdG_tbm}
 \end{align}
where $\mu$ is the chemical potential and $\Delta = \Delta_0 (-i\sigma_0 s_y)$ is an $s$-wave pair potential. 
We then introduce a vortex line along the $z$ axis with the vortex center located at the origin of the $xy$ plane. The pair potential is written in the cylindrical coordinate $(\rho,\theta)$ as
 \begin{align}
  \Delta_0 \to \Delta_0(\rho)e^{i\theta},
 \end{align}
 where the amplitude of the pair potential around the vortex core is approximated by the hyperbolic tangent,
  \begin{equation}
 \Delta_0(\rho) = \Delta_0 \tanh \left(\frac{\rho}{\xi}\right), \label{eq:vortexline}
 \end{equation}
 with the coherence length $\xi$.
 
In the presence of the vortex line, the BdG Hamiltonian is invariant under the following modified screw and inversion operations:
 \begin{align}
&\widetilde{\mathcal{S}}_{4z}^{k_z} = \begin{pmatrix} e^{i \frac{\pi}{4}} \mathcal{S}_{4z}^{k_z} & 0 \\ 0 & e^{-i \frac{\pi}{4}} (\mathcal{S}_{4z}^{-k_z })^{\ast}\end{pmatrix}, \label{eq:screw_tbm_bdg} \\ 
&\widetilde{I}^{k_z} = \begin{pmatrix} e^{i \frac{\pi}{2}} I^{k_z} & 0 \\ 0 & e^{-i \frac{\pi}{2}} (I^{-k_z})^{\ast}\end{pmatrix}. 
\end{align}
On the other hand, $\tilde{C}_{2x}^{k_z} = \diag(C_{2x}^{k_z},C_{2x}^{-k_z \ast})$ is not a symmetry of the BdG Hamiltonian because the two-fold rotation reverses the winding of the vortex. However, the combination of $\widetilde{C}_{2x}^{k_z}$ and $\widetilde{T}= \diag(T,T^{\ast})$ keeps the BdG Hamiltonian invariant at arbitrary wave number $k_z$. The combined $\widetilde{T}\widetilde{C}_{2x}^{k_z}$ symmetry plays a role of time-reversal symmetry with $(\widetilde{T}\widetilde{C}_{2x}^{k_z})^2=\bm{1}_4$, which does not cause additional band degeneracy.
The vortex modes can be described by the low-energy effective Hamiltonian discussed in Sec.~\ref{sec:classification}; see Appendix~\ref{app:k-th} for the classification of vortex line modes including the $\widetilde{T}\widetilde{C}_2$ symmetry.
 
\begin{figure}[tbp]
\centering
 \includegraphics[width=8cm]{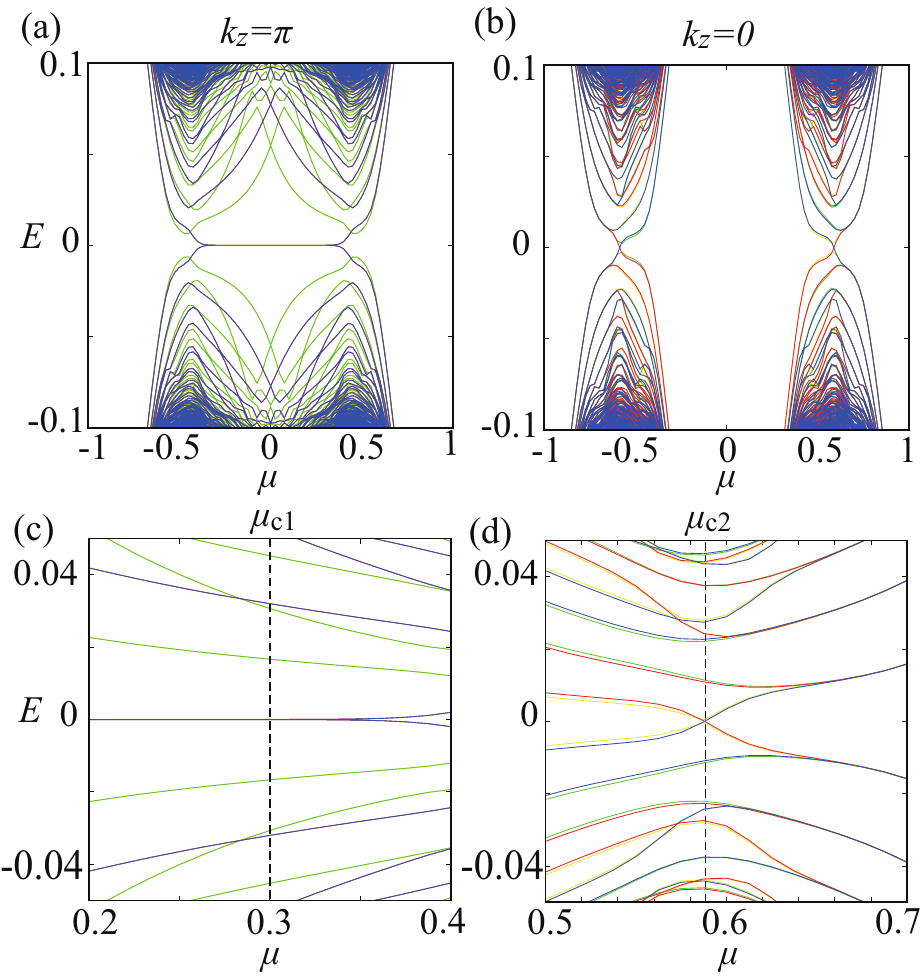}
 \caption{ (Color online) The evolution of vortex bound states as a function of the chemical potential at (a),(c) $k_z=\pi$ and (b),(d) $k_z=0$. (c) and (d) are the enlarged view of (a) and (b) around critical values. The hopping parameters and the pair potential are chosen as in Fig.~\ref{fig:tbvm}. The colors indicating the screw eigenvalues of the energy spectra are the same as those in Fig.~\ref{fig:tbvm}.
 }
\label{fig:tbvm_mu}
\end{figure}
 
The energy spectra of vortex bound states are obtained by numerically diagonalizing the BdG Hamiltonian with the open boundary conditions for the $x$ and $y$ directions and the periodic boundary condition for the $z$ direction. The energy levels are labeled by the $k_z$-dependent screw eigenvalues since the BdG Hamiltonian commutes with the screw operator [Eq.~(\ref{eq:screw_tbm_bdg})]. Here we regard $(\rho,\theta)$ as internal indices, and the screw operator is supplemented by a unitary transformation that exchanges the site indices in the $xy$ plane according to the $\pi/4$ rotation.
 
In Fig.~\ref{fig:tbvm}, we illustrate the evolution of the energy bands of vortex bound states as $\mu$ changes from $0$ to $-0.7$.
At $\mu=0$ [Fig.~\ref{fig:tbvm} (a)], the energy bands are two-fold degenerate.
In particular, there are two-fold degenerate gapless helical modes crossing at $k_z=\pm\pi$.
These gapless modes originate from the Dirac cones in the normal-state Hamiltonian $H(\bm{k})$ at $Z=(0,0,\pi)$ and $A=(\pi,\pi,\pi)$. When coupled with the $s$-wave pair potential, each Dirac cone is described by the Jackiw-Rossi model \cite{JackiwRossi1981} doubled by the sublattice degrees of freedom. For each sublattice and valley ($Z$ and $A$), the Jackiw-Rossi model has a  Majorana zero mode, which generates a total of four Majorana zero modes at $k_z=\pi$. The explicit form of the zero-energy wave functions in the low-energy effective BdG Hamiltonian is presented in Appendix~\ref{app:low-h}.
The two-fold degeneracy of the energy bands at $\mu=0$ can be understood from the following relation valid at $\mu=0$:
\begin{equation}
s_z \sigma_z \tau_0 \widetilde{H}(k_x+\pi,k_x+\pi,k_z) s_z \sigma_z \tau_0 
= - \widetilde{H}(k_x,k_y,k_z) .
\end{equation}

For small but finite $|\mu|$, the energy bands of vortex bound states are not two-fold degenerate, except at the zone boundaries $k_z=\pm\pi$.
Indeed, the four-fold degenerate Majorana zero modes remain at $k_z=\pm\pi$ [Fig.~\ref{fig:tbvm} (b)-(d)].
When $\mu$ changes further ($ \mu \lesssim -0.3$), the four-fold degenerate zero modes are split into a pair of two-fold degenerate gapless points apart from $k_z=\pi$. The stability of these gapless points are ensured by the screw symmetry since they have the different eigenvalues $\pm ie^{-ik_z/2}$ of the screw operator. For $ \mu \lesssim -0.6$, we obtain fully-gapped energy spectra since a pair of two-fold degenerate gapless points meet and annihilate at $k_z=0$.

Figure~\ref{fig:tbvm_mu} shows the evolution of the vortex bound states at $k_z=0$ or $\pi$ as a function of the chemical potential. We notice three characteristic regimes in the behavior of vortex bound states: (I) $|\mu| \le \mu_{\rm c 1}$, (II) $\mu_{\rm c 1} < |\mu|  \le \mu_{\rm c 2}$, (III) $\mu_{\rm c 2} < |\mu|$, where $\mu_{\rm c1}\approx0.3$ and $\mu_{\rm c2}\approx0.58$. In (I), we find the Jackiw-Rossi-type zero-energy modes with four-fold degeneracy at $k_z=\pi$, which remain stable as long as the approximation of the low-energy Hamiltonian to the Dirac Hamiltonian is justified.
In (II), the four-fold degenerate zero modes are split into a pair of two-fold degenerate zero modes away from $k_z=\pi$, and therefore a gap opens at $k_z=\pi$. As $|\mu|$ is increased, the two gapless modes move along the $k_z$ axis [Fig.~\ref{fig:tbvm} (e) and (f)] until they meet and annihilate at $k_z=0$ when $|\mu|=\mu_{\rm c2}$. The trivial phase (III) appears at $|\mu|>\mu_{\rm c2}$. Note that $\mu_{\rm c2}$ signals a topological phase transition since it accompanies the pair-annihilation of zero modes, whereas the transition between the regimes (I) and (II) is considered to be a smooth crossover. It is clear from Eq.~(\ref{eq:screw_kz2}) that the four-fold degeneracy at $k_z=\pi$ is obtained by tuning the parameter $m_0$ to zero.  Since there is no symmetry enforcing $m_0=0$, we expect that the parameter $m_0$ representing small deviations from the Jackiw-Rossi model can be finite in (I).

We comment on the dependence on the position of the vortex core. The results presented above were obtained for the case when the vortex core is positioned at a lattice site. When it is located at the center of a plaquette, we have found that the four-fold-screw-symmetry-protected gapless modes disappear and a fully-gapped phase is realized due to the changes in the screw eigenvalues of low-energy levels. This is because the four-fold screw eigenvalues of the zero-energy modes associated with the Dirac cones depend on the position of the vortex core. The detailed discussion is given in Appendix~\ref{app:VCP}. On the other hand, the stability of the six-fold-screw-symmetry-protected GVL phases is independent of the vortex core position since its mechanism is the EBC.

\subsection{Application to Nb$_3$Pt}
\label{sec:material}

\begin{figure*}[tp]
	\centering
	\includegraphics[width=16cm]{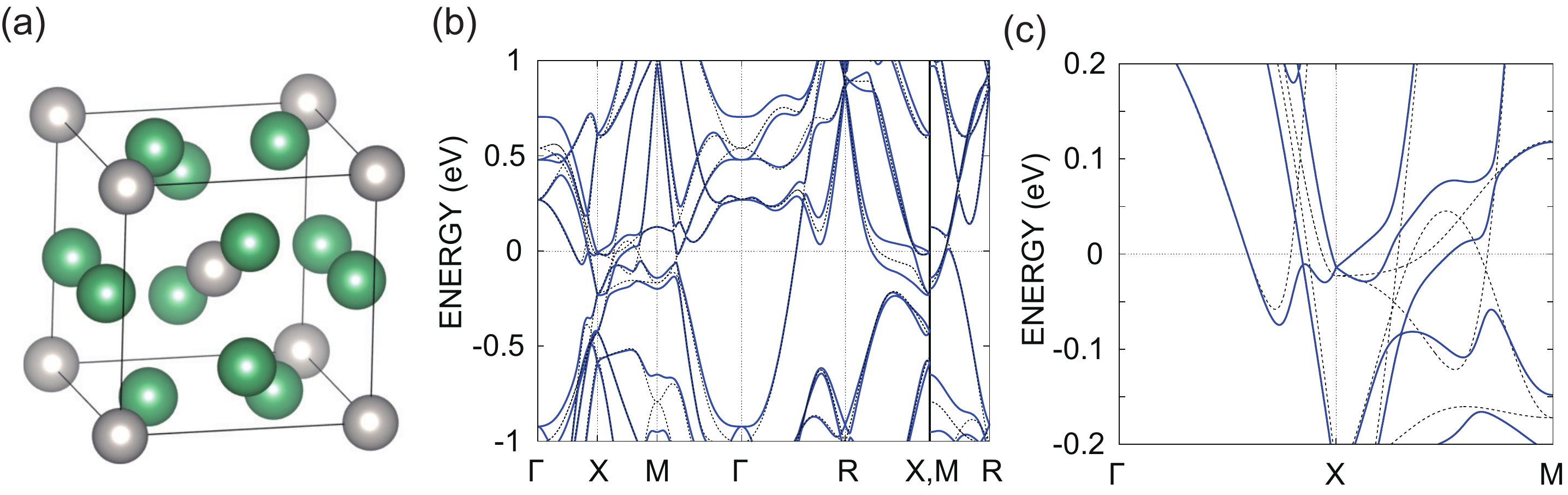}
	\caption{ (Color online) (a) Crystal structure of the Nb$_3$X-family. (b) Electronic band structure of Nb$_3$Pt. (c) The magnified band structure around the X point. The solid and dotted lines correspond to the bands with and without the spin orbit coupling, respectively. The energy is measured from the Fermi level.}
	\label{fig:nb3pt}
\end{figure*}

Finally, we discuss the Nb$_3$X-family as a material candidate that may realize a GVL phase. The crystal structure of the Nb$_3$X-family is shown in Fig.~\ref{fig:nb3pt}(a).
Here, we report the electronic band structures of Nb$_3$Pt obtained from the calculation based on the density functional theory (DFT).
We used the \textit{ab initio} code OpenMX~\cite{PhysRevB.67.155108} with the Perdew-Burke-Ernzerhof (PBE) functional~\cite{PhysRevLett.77.3865} and
the valence orbital sets Nb7.0-s3p3d3f1 and Pt7.0-s2p2d2f1.
The energy cutoff for the numerical integration was set to 150 Ry.
We employed the 12$\times$12$\times$12 $\bm{k}$ mesh.

The crystal structure of the Nb$_3$X-family obeys space group symmetry $Pm\bar{3}n$ (SG\# 223).
The Nb$_3$X-family is a class of materials that exhibit superconductivity with a relatively high upper critical field~\cite{Hughes1975,Flukiger1979,Lambert1980} due to large electron-phonon coupling~\cite{Shen1972,Fisk1976,Knapp1976,Wolf1980}.
Interestingly, the Nb$_3$X-family allows a wide selection of atoms in the X site, so that we can easily tune the position of the Fermi level. 
Substitution of Ta for Nb is also possible, which decreases the transition temperature but enhances the spin-orbit coupling.
Among the Nb$_3$X family, here we focus on Nb$_3$Pt which exhibits superconductivity at $T_{\rm c} =10.9$~\cite{Neuberger1972} since it hosts Dirac points close to the Fermi energy ($\approx-0.02\,$eV) at the $X=(\pi,0,0)$ point and its symmetry related points.
The energy band structures of Nb$_3$Pt along high-symmetry lines are shown in Fig.~\ref{fig:nb3pt} (b) and (c).
All the bands are two-fold degenerate due to the TR and inversion symmetries.
The Dirac point at the $X$ point corresponds to the one at the $Z$ point protected by the screw symmetry discussed in Sec.~\ref{sec:model}.
There is another type of Dirac point on the $\Gamma$-$X$ line protected by the $C_{2x}$ rotation symmetry. 
From the viewpoint of crystalline symmetry, we expect that Nb$_3$Pt should share the same physics with the tight-binding model discussed in Sec.~\ref{sec:model}.

To see this, we examine the group operations of $Pm\bar{3}n$, which are generated by a four-fold screw $\{C_{4z}|\bm{\tau}_d\}$ with $\bm{\tau}_d=(\frac{1}{2},\frac{1}{2},\frac{1}{2})$, a three-fold rotation around the $[111]$ direction $\{C_{3_{[111]}}|\bm{0}\}$, a two-fold rotation around the $[110]$ direction $\{C_{2_{[110]}}|\bm{\tau}_d\}$, and an inversion operator $\{I|\bm{0}\}$~\footnote{The notation $\{g|\bm{a}\}$ is a Seitz space group symbol with a point group operation $g$ and a translation $\bm{a}$.}. When considering an $s$-wave superconducting state with a vortex line along the $z$ axis, only $\{C_{4z}|\bm{\tau}_d\}$, $\{I|\bm{0}\}$, and the combination of TR and $\{C_{2x}|\bm{0}\}$ remain as the relevant symmetry of the system.  Comparing them with the symmetry operations in Eq.~(\ref{eq:tbmodel}), we notice that the only difference is the expression of the four-fold screw operator; $\{C_{4z}|\bm{\tau}_d\}$ accompanies a half translation along the $x$ and $y$ axes in addition to the $z$ axis. Thus, $\{C_{4z}|\bm{\tau}_z\}$ with $\bm{\tau}_z=(0,0,\frac{1}{2})$ can be met when we choose the position of the vortex core such that the translation in terms of the $x$ and $y$ directions vanishes. Such a translation is given by $\bm{\eta}=(0,-\frac{1}{2},0)$ under which $\{C_{4z}|\bm{\tau}_d\}$ changes to
\begin{align}
\{e|\bm{\eta}\}^{-1} \{C_{4z}|\bm{\tau}_d\} \{e|\bm{\eta}\} = \{C_{4z}|\bm{\tau}_z\}
\end{align}
where $e$ is a unit element and we use the multiplication rule:
\begin{align}
 \{g|\bm{a}_g\}\{h|\bm{a}_{h}\} = \{gh|g\bm{a}_h+\bm{a}_g\}.
\end{align}
As a result, the crystal symmetry that preserves superconducting states with a vortex line along $z$ axis becomes $G = G_{\rm vortex}  + \{T C_{2x}|\bm{0}\} G_{\rm vortex} $ with
\begin{align}
G_{\rm vortex} =& \{e|\bm{0}\} \mathbb{T}_z+ \{C_{4z}|\bm{\tau}_z\} \mathbb{T}_z+\{C_{4z}^{-1}|\bm{\tau}_z\}\mathbb{T}_z \nonumber \\
 & +\{C_{2z}|\bm{0}\}\mathbb{T}_z+ \{I|\bm{0}\}\mathbb{T}_z + \{IC_{4z}|\bm{\tau}_z\} \mathbb{T}_z \nonumber \\
 & +\{IC_{4z}^{-1}|\bm{\tau}_z\}\mathbb{T}_z+\{IC_{2z}|\bm{0}\}\mathbb{T}_z,
\end{align}
which are equivalent to the operations in Sec.~\ref{sec:model}. Here, $\mathbb{T}_z$ is the translation group in the $z$ direction.
In addition, Nb$_3$Ir is also a superconductor with $T_{\rm c}=1.7$.
We can adjust the Dirac point to the Fermi level by slightly substituting Pt for Ir.
Therefore, Nb$_3$Pt is expected to be a good test bed to study the relation between screw-symmetry protected Dirac cones and GVL phases, when other metallic bands become $s$-wave superconducting.

\section{Summary}
\label{sec:summary}

We have studied crystal-symmetry protected GVL phases in superconducting Dirac semimetals.
Under the crystal symmetries that host Dirac semimetals, we classified possible GVL modes using the symmetry analysis of the effective low-energy Hamiltonian and discussed the mechanism of the formation of gapless vortex modes for the cases where the Dirac cones are protected by (a) rotation symmetry or (b) screw symmetry.  In the case (a), the effective Hamiltonian is a $2\times 2$ matrix, and the $n$-fold rotation-symmetry-protected GVL modes appear through the ABC [Fig.~\ref{fig:kp} (a)] when the rotation eigenvalues satisfy $2p/n \notin \mathbb{Z}$. In the case (b), the effective Hamiltonian is enlarged to a $4\times 4$ matrix due to the two-fold degeneracy at $k_z=\pi$.  The $n$-fold screw-symmetry-protected GVL modes are realized through the ABC [Fig.~\ref{fig:kp} (b)] for $n=4$ and $p/2 \notin \mathbb{Z}$ and the EBC [Fig.~\ref{fig:kp} (c)] for $n=6$ and $p/3 \notin \mathbb{Z}$.
The $n=6$ EBC case is particularly interesting in that the existence of gapless modes is determined solely by the representation of the screw symmetry, independently of model parameters and the location of the vortex core.

Furthermore, we have demonstrated the four-fold screw-symmetry protected GVL modes in the tight binding model with the space group symmetry $P4_2/mmc$. The model hosts two Dirac cones located at the $Z$ and $A$ points. We considered the $s$-wave superconducting state with a vortex line and found that the vortex bound-state spectra are classified into three regimes as a function of the chemical potential: (I) the Jackiw-Rossi-type zero-energy modes with four-fold degeneracy ($|\mu| < \mu_{\rm c1}$), (II) the four-fold screw-symmetry-protected GVL phase ($\mu_{\rm c1} < |\mu| < \mu_{\rm c2}$), and (III) the fully-gapped vortex-line phase ($\mu_{\rm c2} < |\mu|$).

Finally, we have proposed Nb$_3$Pt as a candidate material to realize screw-symmetry-protected GVL phases. From the viewpoint of crystal symmetry, we have pointed out that the material has both the Dirac cones close the Fermi level in its normal-state band structure and the crystal symmetry that protects screw-symmetry-protected GVL modes when the material is in an $s$-wave superconducting state.

\section*{acknowledgement}
 This work was supported by JSPS KAKENHI (Grants No.\ JP19K03680, No.\ JP19K14612, and No.\ JP19H01824) and JST CREST (Grants No.\ JPMJCR19T2).

\appendix
\section{K-theoretical approach}
\label{app:k-th}

Here, we discuss the K-theoretical classification of GVL phases. As discussed in Sec.~\ref{sec:basis_principle}, the 3D BdG Hamiltonian with a vortex line is mapped to a 1D BdG Hamiltonian. That is, we are able to apply the classification methods developed in the field of nodal superconductors~\cite{Kobayashi14,Kobayashi15PRB,Kobayashi16,tamura2017,Bzdusek2017,Kobayashi18,Sumita19} to the vortex systems. Note that the classification method is also useful to determine electromagnetic response of Majorana zero modes~\cite{Shiozaki14,Xiong2017,Kobayashi19,Yamazaki2019,Yamazaki2021prb,Yamazaki2021jpsj,Kobayashi2021}.  With this in mind, we consider a 1D system with rotation (screw) symmetry $\hat{C}_n \equiv \{C_{nz}|\bm{0}\}$ $(\hat{S}_n \equiv \{C_{nz}|\bm{\tau}_z\})$, inversion symmetry $\hat{I} \equiv \{I | \bm{0}\}$, PH symmetry $\hat{C} \equiv \{C|\bm{0}\}$, and pseudo-TR symmetry $\hat{T}' \equiv \{TC_{2x}|\bm{0} \}$ as the symmetries that are preserved in the presence of a vortex line. We assume that both the rotation (screw) axis and the vortex line are parallel to the $z$ axis. $\hat{C}$ and $\hat{T}'$ are the anti-unitary operators. In the following, we employ a method to classify stable point nodes on a high-symmetry line~\cite{Sumita19}, which can be carried out by determining a local zero-dimensional (0D) topological number from an effective Altland-Zirnbauer (AZ) symmetry class derived from the Wigner's tests in terms of TR, PH, and chiral symmetries.  

First of all, we define $G^{k_z}$ as a little cogroup at $k_z$, which is a set of point group operations that keep $k_z$ invariant up to the reciprocal lattice vector. Full symmetry operations are given by, depending on whether $k_z = k_{\rm inv} = 0,\pi$ or not,
\begin{align}
 G^{k_z} = \left\{\begin{array}{ll}
G_0^{k_{\rm inv}}+ C G_0^{k_{\rm inv}}+T'G_0^{k_{\rm inv}}+CT'G_0^{k_{\rm inv}}, \\
G_0^{k_z} + \hat{\mathfrak{C}} G_0^{k_z}+T'G_0^{k_z}+\hat{\mathfrak{C}} T'G_0^{k_z},  \end{array}\right.
\end{align}
where $\hat{\mathfrak{C}} \equiv \hat{C}\hat{I}$ and $G_0^{k_z}$ is a unitary part of $G^{k_z}$, i.e., $G_0^{k_{\rm inv}} = \{\hat{C}_n, \hat{I}\}$ and $G_0^{k_z} = \{\hat{C}_n\}$ for a rotation-symmetric system; $G_0^{k_{\rm inv}} =\{\hat{S}_n, \hat{I}\}$ and $G_0^{k_z}=\{\hat{S}_n\}$ for a screw-symmetric system. Here, the curly bracket $\{\cdots\}$ represents the generators of the group. 

To formulate the Wigner's test, we introduce $\gamma^{k_z}_{\alpha} (\hat{g})$ as a double-valued (spinful) irrep $\alpha$ of $\hat{g} \in G_0^{k_z}$. The irreps satisfy the multiplication rule:
\begin{align}
z_{\hat{g},\hat{h}}^{\hat{g}\hat{h} k_z} \gamma_{\alpha}^{k_z}(\hat{g}\hat{h})= \begin{cases} \gamma_{\alpha}^{hk_z}(\hat{g}) \gamma_{\alpha}^{k_z} (\hat{h}), &\text{if }\hat{g}\text{ is unitary},  \\ \gamma_{\alpha}^{hk_z}(\hat{g}) \gamma_{\alpha}^{k_z \ast} (\hat{h}), &\text{if }\hat{g}\text{ is antiunitary}, \end{cases} \label{eq:factor}
\end{align}
where $z_{\hat{g},\hat{h}}^{k_z} \in U(1)$ is a so-called factor system. In addition, in the presence of a vortex line, the gauge transformation associated with the vortex field is applied to the irreps. According to Eqs.~(\ref{eq:rs-vortex}) and (\ref{eq:is-vortex}), the irreps are transformed, under $\theta \to \theta +\Delta \theta_g$, as
\begin{align}
 \tilde{\gamma}^{k_z}_{\alpha} (\hat{g}) \equiv e^{i\Delta \theta_g/2}\gamma^{k_z}_{\alpha} (\hat{g}).
\end{align}  
Accordingly, Eq.~(\ref{eq:factor}) is also modified as
\begin{align}
\tilde{z}_{\hat{g},\hat{h}}^{\hat{g}\hat{h} k_z} \tilde{\gamma}_{\alpha}^{k_z}(\hat{g}\hat{h})= \begin{cases} \tilde{\gamma}_{\alpha}^{hk_z}(\hat{g}) \tilde{\gamma}_{\alpha}^{k_z} (\hat{h}), &\text{if }\hat{g}\text{ is unitary},  \\ \tilde{\gamma}_{\alpha}^{hk_z}(\hat{g}) \tilde{\gamma}_{\alpha}^{k_z \ast} (\hat{h}), &\text{if }\hat{g}\text{ is antiunitary}, \end{cases}\label{eq:factor2}
\end{align}
where $\tilde{z}_{\hat{g},\hat{h}}^{k_z} \in U(1)$ includes the vortex-field-indued phase factor.

We employ the Wigner's tests to determine effective AZ classes in terms of the PH ($\hat{C}$ or $\hat{\mathfrak{C}}$), TR ($\hat{T}'$), chiral ($\hat{\Gamma} \equiv \hat{C}\hat{T}'$ or $\hat{\mathfrak{C}}\hat{T}'$) operators. They are explicitly formulated as~\cite{Wigner59,Bradley72,Inui1990} 
\begin{align}
 W_{\alpha}^{C} \equiv& \frac{1}{|G_0^{k_z}|} \sum_{\hat{g} \in G_0^{k_z}} \tilde{z}^{k_z}_{\hat{C}\hat{g},\hat{C}\hat{g}} \tr \{\tilde{\gamma}_{\alpha}^{k_z}[(\hat{C}\hat{g})^2]\} = \pm1 ,0,  \label{eq:W_C} \\
  W_{\alpha}^{T} \equiv& \frac{1}{|G_0^{k_z}|} \sum_{\hat{g} \in G_0^{k_z}} \tilde{z}^{k_z}_{\hat{T}'\hat{g},\hat{T}'\hat{g}}
   \tr \{\tilde{\gamma}^{k_z}_{\alpha}[(\hat{T}'\hat{g})^2]\} = \pm1 ,0,  \label{eq:W_T} \\
  W_{\alpha}^{\Gamma} \equiv& \frac{1}{|G_0^{k_z}|} \sum_{\hat{g} \in G_0^{k_z}} \frac{\tilde{z}^{k_z}_{\hat{g},\hat{\Gamma}}}{\tilde{z}^{k_z}_{\hat{\Gamma},\hat{\Gamma}^{-1}\hat{g}\hat{\Gamma}}}
 \tr [\tilde{\gamma}^{k_z}_{\alpha}(\hat{\Gamma}^{-1}\hat{g}\hat{\Gamma})]^{\ast} \tr [ \tilde{\gamma}^{k_z}_{\alpha}(\hat{g})]
 \nonumber\\ &
  = 1,0, \label{eq:W_G}
\end{align}
when $k_z = k_{\rm inv}$, whereas $\hat{C}$ is replaced with $\hat{\mathfrak{C}}$ when $k_z \neq k_{\rm inv}$. Here, $\tilde{z}_{\hat{C},\hat{C}}=\tilde{z}_{\hat{T}',\hat{T}'}=1$, and $\tilde{z}_{\hat{C},\hat{g}}=\tilde{z}_{\hat{g},\hat{C}}=1$ since we consider $s$-wave SCs. The triples $(W_{\alpha}^{C} ,W_{\alpha}^{T} ,W_{\alpha}^{\Gamma})$ determine the effective AZ classes and the corresponding 0D topological numbers as shown in Table~\ref{tab:AZclass}.

\begin{table}[tb]
\caption{
The effective AZ symmetry classes and the 0D topological numbers for $k_z = k_{\rm inv}$. Here, $C$ is replaced by $\mathfrak{C}$ for $k_z \neq k_{\rm inv}$.
}
\label{tab:AZclass}
\begin{tabular}{ccccc} \hline \hline 
AZ class & $W_{\alpha}^T$ & $W_{\alpha}^C$ & $W_{\alpha}^{\Gamma}$ & 0D topo. \# \\
\hline 
A & $0$& $0$& $0$& $\mathbb{Z}$ \\
AIII & $0$& $0$& $1$& $0$ \\
AI &  $1$& $0$& $0$& $\mathbb{Z}$ \\
BDI & $1$& $1$& $1$& $\mathbb{Z}_2$ \\
D &  $0$& $1$& $0$& $\mathbb{Z}_2$ \\
DIII &  $-1$& $1$& $1$& $0$ \\
AII &  $-1$& $0$& $0$& $2\mathbb{Z}$ \\
CII &  $-1$& $-1$& $1$& $0$ \\
C & $0$& $-1$& $0$& $0$ \\
CI &  $1$& $-1$& $1$& $0$ \\
\hline\hline
\end{tabular} 
\end{table} 

Let us perform the Wigner's tests for a system with four-fold screw and inversion symmetries, i.e., $G_0^{k_{\rm inv}} = \{\hat{S}_{4},\hat{I}\}$ and $G_0^{k_z} = \{\hat{S}_{4}\}$, as an example. In this case, the irreps are given by two 2D irreps, 
\begin{subequations}
\label{eq:irrep_pi}
\begin{align}
&\tilde{\gamma}_1^{\pi}(\hat{S}_4) = \begin{pmatrix} 0 & -e^{-i\frac{\pi}{4}} \\ e^{i\frac{\pi}{4}} & 0\end{pmatrix}, \quad \tilde{\gamma}_1^{\pi}(\hat{I}) = \begin{pmatrix} i & 0 \\ 0 & -i\end{pmatrix}, \label{eq:irrep1_pi}\\
&\gamma_2^{\pi}(\hat{S}_4) = \begin{pmatrix} 0 & e^{-i\frac{\pi}{4}} \\ e^{i\frac{\pi}{4}} & 0\end{pmatrix}, \quad 
\tilde{\gamma}_2^{\pi}(\hat{I}) = \begin{pmatrix} i & 0 \\ 0 & -i\end{pmatrix}. \label{eq:irrep2_pi}
\end{align}
\end{subequations}
at $k_z =\pi$, four 1D irreps,
\begin{subequations}
\label{eq:irrep_kz} 
\begin{align}
&\tilde{\gamma}_{1\pm}^{k_z}(\hat{S}_4) = \pm e^{-ik_z/2} \label{eq:irrep1_kz} \\
&\tilde{\gamma}_{2\pm}^{k_z}(\hat{S}_4) = \pm i e^{-ik_z/2}, \label{eq:irrep2_kz}
\end{align}
\end{subequations}
at $k_z \neq 0,\pi$, and eight 1D irreps,
\begin{subequations}
\label{eq:irrep_0} 
\begin{align}
&\tilde{\gamma}_{1\pm}^0(\hat{S}_4) = \pm 1, \quad \tilde{\gamma}_{1\pm}^0(\hat{I}) = i, \label{eq:irrep1_0} \\
&\tilde{\gamma}_{1'\pm}^0(\hat{S}_4) = \pm 1, \quad \tilde{\gamma}_{1'\pm}^0(\hat{I}) = -i, \label{eq:irrep1p_0} \\
&\tilde{\gamma}_{2\pm}^0(\hat{S}_4) = \pm i, \quad \tilde{\gamma}_{2\pm}^0(\hat{I}) = i,  \label{eq:irrep2_0} \\
&\gamma_{2'\pm}^0(\hat{S}_4) = \pm i, \quad \tilde{\gamma}_{2'\pm}^0(\hat{I}) = -i, \label{eq:irrep2p_0} 
\end{align}
\end{subequations}
at $k_z=0$, where we have used the irreps in the Bilbao Crystallographic Server~\cite{Elcoro2017}, $P4_2/m$ (SG\# 84), and the vortex-field-induced additional phases are $\Delta \theta_{S_4} = \pi/4$ and $\Delta \theta_{I} = \pi/2$.
The compatibility relation imposes the constraints
\begin{align}
 &\tilde{\gamma}^{\pi}_\ell \downarrow G^{k_z}_0 =  \tilde{\gamma}^{k_z}_{\ell+} + \tilde{\gamma}^{k_z}_{\ell-}, \\ 
 &\tilde{\gamma}^{0}_{\ell \pm} \downarrow G^{k_z}_0 =  \tilde{\gamma}^{k_z}_{\ell\pm}, \\
 &\tilde{\gamma}^{0}_{\ell' \pm} \downarrow G^{k_z}_0 =  \tilde{\gamma}^{k_z}_{\ell\pm}, 
\end{align}
where $\ell=1,2$ and $\tilde{\gamma} \downarrow G^{k_z}_0$ means the decomposition of  $\tilde{\gamma}$ to irreps on $G^{k_z}_0 \subset G^{k_{\rm inv}}_0$.
Substituting Eqs.~(\ref{eq:irrep_pi}), (\ref{eq:irrep_kz}), (\ref{eq:irrep_0}) into the Wigner's tests (\ref{eq:W_C}), (\ref{eq:W_T}), and (\ref{eq:W_G}), we obtain
\begin{align}
 &W_{\ell}^C = 
 \begin{cases} 
 -1 & \text{when }\ell = 1, \\
 1 & \text{when }\ell = 2, 
 \end{cases} \\
 &W_{\ell}^T = 1 \ \ \text{for }^{\forall}\ell, \\
 &W_{\ell}^\Gamma = 1 \ \ \text{for }^{\forall}\ell,
\end{align}
at $k_z =\pi$,
\begin{align}
 &W_{\ell \pm}^{\mathfrak{C}} = 
 \begin{cases} 
 -1 & \text{when }\ell = 1, \\
 0 & \text{when }\ell = 2,
 \end{cases} \\
 &W_{\ell \pm}^T = 1 \ \ \text{for }^{\forall}\ell, \\
 &W_{\ell \pm}^\Gamma =  
  \begin{cases} 
 1 & \text{when }\ell = 1, \\
 0 & \text{when }\ell = 2,
 \end{cases}
\end{align}
at $k_z \neq 0,\pi$, and
\begin{align}
 &W_{\ell \pm}^C = 
0 \ \ \text{for }^{\forall}\ell, \\
 &W_{\ell \pm}^T = 0 \ \ \text{for }^{\forall}\ell, \\
 &W_{\ell \pm}^\Gamma =  
  \begin{cases} 
 1 & \text{when }\ell = 1,1' , \\
 0 & \text{when }\ell = 2,2' ,
 \end{cases}
\end{align}
at $k_z=0$. The obtained results mean that $\tilde{\gamma}_1^{\pi}$ and $\tilde{\gamma}_2^{\pi}$ belong to CI and BDI in the AZ class; $\tilde{\gamma}_{1 \pm}^{k_z} $ and $\tilde{\gamma}_{2 \pm}^{k_z} $ to CI and A; $\tilde{\gamma}_{1 \pm}^0 $, $\tilde{\gamma}_{1' \pm}^0 $, $\tilde{\gamma}_{2 \pm}^0 $ $\tilde{\gamma}_{2' \pm}^0 $ to AIII, AIII, A, and A, respectively. Combining these results with the compatibility constraints and determining the corresponding 0D topological number from the AZ class (see Table~\ref{tab:AZclass}), we find two types of node structures in terms of $k_z$:
\begin{equation}
\begin{array}{c|cccccc}
{\rm irrep} & k_z=\pi &\to& k_z\neq 0,\pi&\to & k_z=0 \\ \hline 
\tilde{\gamma}_{1}^{\pi}(\hat{S}_4) & 0 && 0 && 0 \\
\tilde{\gamma}_{2}^{\pi}(\hat{S}_4) &\mathbb{Z}_2 && \mathbb{Z} && \mathbb{Z}
\end{array}  \label{eq:vortex_sq}
\end{equation}
where the leftmost column indicates the irreps at $k_z=\pi$ and the other columns represent the 0D topological numbers for each $k_z$.
Equation~(\ref{eq:vortex_sq}) implies that the GVL phases are possible for $\tilde{\gamma}_{2}^{\pi}(\hat{S}_4)$ because we have nontrivial 0D topological numbers classified by $\mathbb{Z}$ at $k_z\neq 0,\pi$, which stabilizes a band crossing at arbitrary $k_z$. Interestingly, we find the different topological numbers at $k_z=\pi$, which may indicate types of the underlying mechanisms for the formation of GVL modes. For comparison, we show the topological classification in the case of six-fold screw symmetry, which are obtained as
\begin{equation}
\begin{array}{c|cccccc}
{\rm irrep} & k_z=\pi &\to& k_z\neq 0,\pi&\to & k_z=0 \\ \hline 
\tilde{\gamma}_{1}^{\pi}(\hat{S}_6) &\mathbb{Z} && \mathbb{Z} && \mathbb{Z} \\
\tilde{\gamma}_{2}^{\pi}(\hat{S}_6)  & 0 && 0 && 0 \\
\tilde{\gamma}_{3}^{\pi}(\hat{S}_6) &\mathbb{Z} && \mathbb{Z} && \mathbb{Z}
\end{array},  \label{eq:vortex_sq6}
\end{equation}
where the irreps are defined by, using the the Bilbao Crystallographic Server~\cite{Elcoro2017}, $P6_3/m$ (SG\# 176),
\begin{subequations}
\begin{align}
&\tilde{\gamma}_1^{\pi}(\hat{S}_6) = \begin{pmatrix} 0 & e^{i\frac{\pi}{6}} \\ e^{i\frac{\pi}{6}} & 0\end{pmatrix}, \quad \tilde{\gamma}_1^{\pi}(\hat{I}) = \begin{pmatrix} i & 0 \\ 0 & -i\end{pmatrix}, \label{eq:irrep61_pi} \\
&\gamma_2^{\pi}(\hat{S}_6) = \begin{pmatrix} 0 & -i \\ -i & 0\end{pmatrix}, \quad 
\tilde{\gamma}_2^{\pi}(\hat{I}) = \begin{pmatrix} i & 0 \\ 0 & -i\end{pmatrix}, \label{eq:irrep62_pi} \\
&\tilde{\gamma}_3^{\pi}(\hat{S}_6) = \begin{pmatrix} 0 & e^{i\frac{5\pi}{6}} \\ e^{i\frac{5\pi}{6}} & 0\end{pmatrix}, \quad \tilde{\gamma}_3^{\pi}(\hat{I}) = \begin{pmatrix} i & 0 \\ 0 & -i\end{pmatrix}, \label{eq:irrep63_pi}
\end{align}
\end{subequations}
with $\Delta \theta_{S_6}=\pi/6$ and $\Delta \theta_{I}=\pi/2$. Thus, stable GVL phases appear when the irreps at $k_z=\pi$ are given by $\tilde{\gamma}_{1}^{\pi}(\hat{S}_6)$ and $\tilde{\gamma}_{3}^{\pi}(\hat{S}_6)$, and the 0D topological number at $k_z=\pi$ is different from that in Eq.~(\ref{eq:vortex_sq}). In view of this, Eqs.~(\ref{eq:vortex_sq}) and (\ref{eq:vortex_sq6}) are consistent with the symmetry analysis of the effective low-energy Hamiltonians in Sec.~\ref{sec:basis_principle}.

\begin{table}[tb]
\caption{
The topological classification of nodal phases in a 1D superconducting vortex line with $n$-fold rotation symmetry. The symmetries of the systems are the $n$-fold rotation symmetry $\{C_{nz}|\bm{0}\}$, inversion symmetry $\{I | \bm{0}\}$, PH symmetry $\{C|\bm{0}\}$, and pseudo-TR symmetry $\{TC_{2x}|\bm{0} \}$. The first and second columns show the irreps of $n$-fold rotation operator~(\ref{eq:rot_irrep}). The other columns represent the AZ classes at $k_z= k_{\rm inv} = 0, \pi$ and $k_z \neq k_{\rm inv}$, where the numbers in the parentheses are the corresponding 0D topological numbers and $\checkmark $ is marked for GVL phases.
}
\label{tab:rot_EAZclass}
\begin{tabular}{ccccc}
\hline\hline
$n$ & $p$ & $k_z= k_{\rm inv}$ &  $k_z \neq k_{\rm inv}$ &GVL modes\\
\hline 
$2$ & 1,2  & AIII (0) & CI (0) &  \\
$3$ & 3   &  AIII (0) & CI (0)  &  \\
$3$ & 1,2 &  A ($\mathbb{Z}$) & AI ($\mathbb{Z}$)& \checkmark  \\
$4$ & 2,4 &  AIII (0) & CI (0)  &  \\
$4$ & 1,3 &  A ($\mathbb{Z}$) & AI($\mathbb{Z}$)& \checkmark \\
$6$ & 3,6 &  AIII (0) & CI (0)  &  \\
$6$ & 1,2,4,5 &  A ($\mathbb{Z}$) &AI ($\mathbb{Z}$)& \checkmark \\
\hline\hline
\end{tabular} 
\end{table} 

\begin{table}[tb]
\caption{
The topological classification of nodal phases in the case of $n$-fold screw symmetry $\{C_{nz}|\bm{\tau}_z\}$. The first and second columns represent the irreps of the screw operator~(\ref{eq:screw_irrep}).
 In third and forth columns, we show the AZ symmetry classes and the $0$D topological numbers at $k_z= \pi$ and $k_z \neq k_{\rm inv}$. The classification at $k_z=0$ is the same as the third column in Table~\ref{tab:rot_EAZclass}.
}
\label{tab:screw_EAZclass}
\begin{tabular}{ccccc}
\hline\hline
$n$ & $p$ & $k_z= \pi$ &  $k_z \neq k_{\rm inv}$ &GVL modes\\
\hline 
$2$ & 1  & CI (0) & CI (0) &  \\
$4$ & 2 &  CI (0) & CI (0)  &  \\
$4$ & 1 &  BDI ($\mathbb{Z}_2$) & AI ($\mathbb{Z}$)& \checkmark \\
$6$ & 3 &  CI (0) & CI (0)  &  \\
$6$ & 1,2 &  AI ($\mathbb{Z}$) &AI ($\mathbb{Z}$)& \checkmark \\
\hline\hline
\end{tabular} 
\end{table} 

The topological classifications of nodal phases in a 1D super conducting vortex line with $n$-fold rotation and $n$-fold screw symmetries are summarized in Tables~\ref{tab:rot_EAZclass} and~\ref{tab:screw_EAZclass}, respectively, where we use the same rotation (screw) operators as in Tables~\ref{tab:rot_class} and~\ref{tab:screw_class} for comparison purposes. We find that the two classifications are consistent with each other. Note that in the four-fold screw symmetry, $\tilde{\gamma}_1^{\pi}(\hat{S}_4)$ and $\tilde{\gamma}_2^{\pi}(\hat{S}_4)$ correspond to $p=2$ and $1$, whereas in the six-fold screw symmetry, $\tilde{\gamma}_1^{\pi}(\hat{S}_6)$, $\tilde{\gamma}_2^{\pi}(\hat{S}_6)$, and $\tilde{\gamma}_3^{\pi}(\hat{S}_6)$ correspond to $p=2$, $3$, and $1$, respectively.

\section{Low-energy Dirac Hamiltonian and vortex-zero-energy modes}
\label{app:low-h}

In this appendix, we derive a low-energy Dirac Hamiltonian of the tight-binding Hamiltonian~(\ref{eq:tbmodel}) around $Z=(0,0,\pi)$ and $A=(\pi,\pi,\pi)$ and an associated vortex-zero-energy mode. By expanding Eq.~(\ref{eq:tbmodel}) around the $Z$ and $A$ points, the low-energy Dirac Hamiltonian is given by, up to the linear order of momentum,
\begin{align}
 H_{\nu} (\bm{k}) = 2 \nu t_1 \bm{1}_4 +\frac{t_z}{2} \tilde{k}_z  \sigma_y s_0
  + \nu \lambda_1 \sigma_x (k_y s_x + k_x s_y)\label{eq:low_Hami}
\end{align}
where $\tilde{k}_z =k_z -\pi$, and $\nu=+1$ $(-1)$ corresponds to the $Z$ ($A$) point.
When $\nu=-1$, the 2D momentum $(k_x,k_y)$ is measured from $(\pi,\pi)$.
Similarly, the symmetry operators are also expanded around $Z$ and $A$, which result in
\begin{align}
 &\mathcal{S}_{4z}^{\pi} = i \sigma_y e^{-i\frac{\pi}{4} s_z}, \\ 
 &C_{2x}^{\pi} = i \sigma_z s_x, \\ 
 &I^{\pi} = \sigma_z s_0. 
\end{align}

The $s$-wave superconducting state of the low-energy Dirac Hamiltonian is described by the BdG Hamiltonian: 
\begin{align}
\widetilde{H}_\nu(\bm{k}) =& 
\begin{pmatrix}
 H_{\nu}(\bm{k}) -\mu & \Delta_0 (-i \sigma_0s_y) \\
 \Delta_0^* i \sigma_0s_y &  -H_{\nu}^T(-\bm{k}) +\mu 
\end{pmatrix}
\nonumber\\
=&\,
(2\nu t_1-\mu)\sigma_0s_0\tau_z+\frac{t_z}{2}\tilde{k}_z\sigma_ys_0\tau_z
\nonumber\\ &
+\nu\lambda_1\sigma_x(k_ys_x\tau_0+k_xs_y\tau_z)
\nonumber\\ &
+\sigma_0s_y[\mathrm{Re}(\Delta_0)\tau_y+\mathrm{Im}(\Delta_0)\tau_x]
\label{eq:low_Hamibdg}
\end{align}
with $\Delta_0$ being the $s$-wave pair potential.
We use the unitary operator $U=\frac{1}{\sqrt{2}}(\sigma_z+\sigma_x)s_0 \tau_0$ to transform the BdG Hamiltonian to the form of the Jackiw-Rossi model,
\begin{align}
U \widetilde{H}_\nu(\bm{k}) U^{\dagger} = 
\begin{pmatrix}
 \widetilde{H}_{\nu,+}(k_x,k_y) & \frac{i}{2}t_z \tilde{k}_z s_0 \tau_z\\
 -\frac{i}{2}t_z \tilde{k}_z s_0 \tau_z & \widetilde{H}_{\nu,-}(k_x,k_y)
 \end{pmatrix}_\sigma \label{eq:low_HamibdgU}
\end{align} 
in the $\sigma$ grading, where
\begin{align}
 \widetilde{H}_{\nu,\pm}(k_x,k_y) =&\, (2 \nu t_1-\mu) s_0 \tau_z
 + s_y [\mathrm{Re}(\Delta_0)\tau_y + \mathrm{Im}(\Delta_0)\tau_x] 
 \nonumber\\
 &\pm \nu \lambda_1 (k_y s_x \tau_0 + k_x s_y \tau_z) . \label{eq:low_Hpm}
\end{align}

We implement a vortex positioned at $x=y=0$ in the pair potential, which is described by $\Delta_0 = \Delta(\rho)e^{i\theta}$, $k_x \to - i \partial_x = -i[\cos (\theta) \partial_\rho - \sin (\theta) \partial_{\theta}/\rho]$, and $k_y \to - i \partial_y = -i[\sin (\theta) \partial_\rho + \cos (\theta) \partial_{\theta}/\rho]$. Here we assume that $\Delta(\rho)$ satisfies $\Delta(\rho=0)=0$ and $\Delta(\rho \to \infty)=\Delta_0$ and use the polar coordinate $(x,y)=(\rho \cos(\theta),\rho \sin(\theta))$. Note that the translation symmetry along the $z$ axis is preserved. Equation (\ref{eq:low_Hpm}) is then rewritten as
\begin{widetext}
\begin{align}
 \widetilde{H}_{\nu,\sigma} (\rho,\theta)= 
 \begin{pmatrix} 
 -m_\nu & -\sigma\nu \lambda_1 e^{i \theta}\left(\partial_\rho +\frac{i}{\rho}\partial_\theta\right) & 0 & -\Delta(\rho)e^{i\theta} \\
 \sigma\nu \lambda_1 e^{-i \theta}\left(\partial_\rho - \frac{i}{\rho}\partial_\theta\right) & -m_\nu & \Delta(\rho)e^{i\theta}  & 0 \\
 0 & \Delta(\rho)e^{-i\theta} & m_\nu & \sigma\nu \lambda_1 e^{-i \theta}\left(\partial_\rho -\frac{i}{\rho}\partial_\theta\right) \\
 -\Delta(\rho)e^{-i\theta} & 0& -\sigma\nu \lambda_1 e^{i \theta}\left(\partial_\rho + \frac{i}{\rho}\partial_\theta\right) &m_\nu
 \end{pmatrix}, \label{eq:low_Hv}
\end{align}
\end{widetext}
where $\sigma =\pm$ and $m_\nu = \mu - 2\nu t_1$.
Equation (\ref{eq:low_Hv}) is equivalent to the Jackiw-Rossi model~\cite{JackiwRossi1981} with the mass $m_\nu$. 
When $k_z=\pi$, the off-diagonal terms in Eq.~(\ref{eq:low_HamibdgU}) vanish. We can find zero-energy solutions by solving Eq.~(\ref{eq:low_Hv}). To see this, consider the eigenvalue problem,  
\begin{align}
U \widetilde{H}_\nu(\bm{k}) U^{\dagger}| \Psi_{\nu,\sigma} (\rho,\theta) \rangle =E |\Psi_{\nu,\sigma} (\rho,\theta) \rangle. \label{eq:diff_equ}
\end{align}
Since Eq.~(\ref{eq:low_HamibdgU}) is block-diagonalized at $\tilde{k}_z=0$, the eigenvalue problems of $\widetilde{H}_{\nu,+}$ and $\widetilde{H}_{\nu,-}$ are solved separately.
Substituting $| \Psi_{\nu,+} (\rho,\theta) \rangle = | \psi_{\nu,+} (\rho,\theta) \rangle \oplus \bm{0}$ and $| \Psi_{\nu,-} (\rho,\theta) \rangle = \bm{0} \oplus | \psi_{\nu,-} (\rho,\theta) \rangle$ into Eq.~(\ref{eq:diff_equ}), the differential equation for the zero-energy states becomes $\widetilde{H}_{\nu,\sigma}(\rho,\theta)| \psi_{\nu,\sigma} (\rho,\theta) \rangle= 0$.  Following Ref.~\onlinecite{JackiwRossi1981,Chamon2010}, we obtain
\begin{align}
 \psi_{\nu,+} (\rho,\theta)  =&\, N_{\nu}
 \exp\!\left(-\int_0^\rho \frac{\Delta(x)}{\lambda_1} dx\right)
  \nonumber\\ & \times
 \left(\begin{array}{@{\,}c@{\,}}
  J_1 \!\left(\frac{m_\nu}{\lambda_1} \rho \right) e^{i\left( \theta+\frac{\pi (\nu-1)}{4}\right)} \\
  J_0 \!\left(\frac{m_\nu}{\lambda_1} \rho\right)e^{-i\frac{\pi (\nu-1)}{4}}  \\
  J_1 \!\left(\frac{m_\nu}{\lambda_1} \rho\right) e^{-i\left( \theta+\frac{\pi (\nu-1)}{4}\right)} \\
  J_0 \!\left(\frac{m_\nu}{\lambda_1} \rho\right)e^{i\frac{\pi (\nu-1)}{4}} \end{array} \right) ,
   \label{eq:VZM1}\\
 \psi_{\nu,-} (\rho,\theta) =&\, N_{\nu} 
 \exp\!\left(-\int_0^\rho \frac{\Delta(x)}{\lambda_1} dx\right)
  \nonumber\\ & \times
 \left(\begin{array}{@{\,}c@{\,}}
  J_1 \!\left(\frac{m_\nu}{\lambda_1} \rho \right) e^{i\left( \theta-\frac{\pi (\nu+1)}{4}\right)} \\
  J_0 \!\left(\frac{m_\nu}{\lambda_1} \rho \right)e^{i\frac{\pi (\nu+1)}{4}} \\
  J_1 \!\left(\frac{m_\nu}{\lambda_1} \rho \right)e^{-i\left( \theta-\frac{\pi (\nu+1)}{4}\right)}  \\
  J_0 \!\left(\frac{m_\nu}{\lambda_1} \rho \right)e^{-i\frac{\pi (\nu+1)}{4}} \end{array} \right)
 , \label{eq:VZM2}
\end{align}
where $J_k(\rho)$ is the $k$th Bessel function and $N_{\nu}$ are normalization constants determined by the condition
\begin{eqnarray}
N_\nu^{-2} &=& 4\pi\int^\infty_0\rho d\rho\exp\!\left(-2\int^\rho_0\frac{\Delta(\rho')}{\lambda_1}d\rho'\right)
\nonumber\\
&&\qquad\times
[J_1^2(m_\nu\rho/\lambda_1) + J_0^2(m_\nu\rho/\lambda_1)].
\end{eqnarray} 
The zero-mode solutions satisfy the Majorana condition
\begin{equation}
|\psi_{\nu,\sigma}\rangle = \tau_x |\psi_{\nu,\sigma}\rangle^*.
\end{equation} 

The off-diagonal components $\pm\frac{i}{2}t_z\tilde{k}_zs_0\tau_z$ in Eq.~(\ref{eq:low_HamibdgU}) mix $|\Psi_{\nu,+} \rangle$ with $|\Psi_{\nu,-} \rangle$ and vice versa, leading to a finite energy proportional to $\tilde{k}_z$.
The matrix elements are calculated as
\begin{subequations}
\begin{eqnarray}
\langle \psi_{\nu,+}|s_0\tau_z|\psi_{\nu,-}\rangle&=&
4\pi i \nu N_\nu^2\int\!\rho d\rho\exp\!\left(-2\!\int^\rho_0\!\frac{\Delta(\rho')}{\lambda_1}d\rho'\right)
\nonumber\\
&&\qquad\times[J_0^2(m_\nu \rho/\lambda_1)-J_1^2(m_\nu \rho/\lambda_1)]
\nonumber\\
&=:&
i\nu c_\nu, \\
\langle \psi_{\nu,-}|s_0\tau_z|\psi_{\nu,+}\rangle&=&
-i\nu c_\nu,
\end{eqnarray}
\end{subequations}
where $c_\nu$ is a constant that depends on $|m_\nu|$.
Then the matrix elements of the BdG Hamiltonian in Eq.\ (\ref{eq:low_HamibdgU}) for $(|\Psi_{\nu,+}\rangle, |\Psi_{\nu,-}\rangle)$ are given by
\begin{align}
\langle  U \widetilde{H}_\nu(\bm{k}) U^{\dagger} \rangle = -\frac{\nu c_\nu t_z}{2} \tilde{k}_z \begin{pmatrix} 0 & 1 \\ 1 & 0 \end{pmatrix}. \label{eq:Hami_eigen}
\end{align}

To check the screw eigenvalues of the zero-energy modes, we evaluate the matrix elements of $\widetilde{\mathcal{S}}_4^{\pi}$ in terms of the zero-energy solutions~(\ref{eq:VZM1}) and (\ref{eq:VZM2}).
From Eq.~(\ref{eq:screw_tbm_bdg}), $\widetilde{\mathcal{S}}_{4z}^{\pi}$ is represented by
\begin{align}
 \widetilde{\mathcal{S}}_{4z}^{\pi} = i \sigma_y e^{- i \frac{\pi}{4} (s_z -s_0)\tau_z},
\end{align}
which is transformed to
\begin{equation}
U\widetilde{\mathcal{S}}_{4z}^{\pi}U^{\dagger} =
 -i \sigma_y e^{-i \frac{\pi}{4}(s_z -s_0) \tau_z}
\end{equation}
by the unitary transformation $U$.
Taking into account the $\pi/2$ rotation in the real space, we calculate the matrix elements
\begin{eqnarray}
&&
\int \!\rho d\rho\int \!d\theta\,[\psi_{\nu,\pm}(\rho,\theta)]^\dagger e^{-i\frac{\pi}{4}(s_z-s_0)\tau_z}\psi_{\nu,\mp}(\rho,\theta-\pi/2)
\nonumber\\
&&\qquad =\mp\nu,
\end{eqnarray}
to write the screw operator $\widetilde{\mathcal{S}}_{4z}^{\pi}$ in the basis set of $(|\Psi_{\nu,+}\rangle, |\Psi_{\nu,-}\rangle)$
\begin{align}
 \langle U \widetilde{\mathcal{S}}_{4z}^{\pi} U^{\dagger} \rangle= + \nu \begin{pmatrix} 0 &1 \\ 1 & 0\end{pmatrix}. \label{eq:S4_eigen}
\end{align}
The linear combinations
\begin{align}
 &| \Psi_{\nu,1}\rangle= \frac{1}{\sqrt{2}}\left( | \Psi_{\nu,+}\rangle+| \Psi_{\nu,-}\rangle \right), \\
 &| \Psi_{\nu,2}\rangle= \frac{1}{\sqrt{2}}\left( | \Psi_{\nu,+}\rangle-| \Psi_{\nu,-}\rangle \right),
\end{align}
diagonalize Eqs.~(\ref{eq:Hami_eigen}) and (\ref{eq:S4_eigen}) simultaneously, yielding
\begin{align}
 &\langle U \widetilde{H}_\nu(\bm{k}) U^{\dagger}\rangle = -\frac{\nu c_\nu t_z}{2} \tilde{k}_z
  \begin{pmatrix} 1 & 0 \\ 0 & -1 \end{pmatrix},
  \label{eq:ene-eigenv1-1} \\
 &\langle U \widetilde{\mathcal{S}}_{4z}^{\pi} U^{\dagger} \rangle = + \nu \begin{pmatrix} 1 & 0 \\ 0 & -1 \end{pmatrix}.
  \label{eq:ene-eigenv1-2}
\end{align} 
The gapless Majorana modes have linear dispersion around $k_z=\pi$ and the screw eigenvalues $\pm1$, which is consistent with the numerical results in Fig.~\ref{fig:tbvm} (a)-(c).

Incidentally, when $\mu=0$, the two masses are related $m_{+1}=-m_{-1}$, and $c_{+1}=c_{-1}$.
Therefore, the $\nu=\pm1$ modes are degenerate.
In fact, all the energy bands of the vortex bound states of the BdG Hamiltonian are two-fold degenerate at $\mu=0$ because $\widetilde{H}_{+1}$ and $\widetilde{H}_{-1}$ have the same energy spectra, which can be understood from the relation
\begin{equation}
 \sigma_x s_0 \tau_z \widetilde{H}_{-1}(\bm{k})\, \sigma_x s_0 \tau_z\left|_{\mu=0}
=-\widetilde{H}_{+1}(\bm{k})\right|_{\mu=0}.
\end{equation}

\section{Influence of vortex core positions on GVL modes}
\label{app:VCP}

\begin{figure}[tbp]
\centering
 \includegraphics[width=8.5cm]{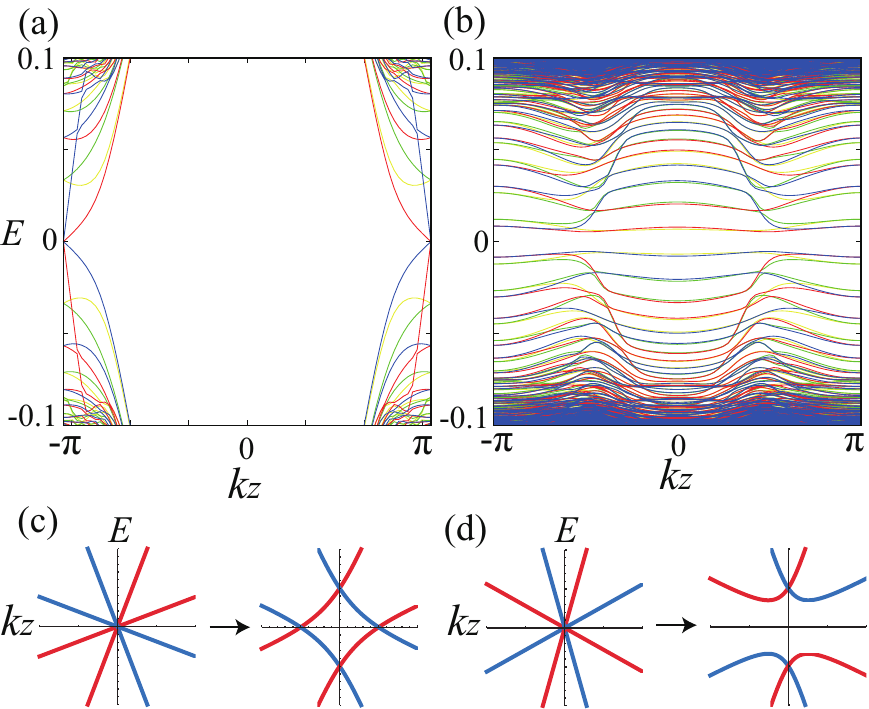}
 \caption{ (Color online) The evolution of vortex bound states as a function of $k_z$ for (a) $\mu = -0.1$ and (b) $-0.5$. The hopping parameters and the pair potential are chosen as the same as in Fig.~\ref{fig:tbvm}. The lattice size is $L=22$, and the vortex core is located at the center of the plaquette. The color code of the energy spectra is the same as in Fig.~\ref{fig:tbvm}.
The panels (c) and (d) show the energy spectrum of the low-energy effective Hamiltonian~(\ref{eq:screw_H2}) around $k_z=\pi$. The arrow between the left and right figures in (c) and (d) indicates the change in the spectra due to the $m_0$ term, which corresponds to changing the chemical potential in the tight-binding model. The parameters are chosen to be $(m_0,v_1,v_2,v_3) = (0,1.5,1,0.5)$ and $(0.5,1.5,1,0.5)$ for the left and right figures of (c); $(0,1.5,2,0.5)$ and $(0.5,1.5,2,0.5)$ for the left and right figures of (d). }
\label{fig:tbvm_even}
\end{figure}

As discussed in Sec.~\ref{sec:classification}, crystal-symmetry-protected GVL modes emerge when vortex bound states have different rotation (screw) eigenvalues. Thus, the stability of GVL modes is sensitive to the representation of rotation (screw) operators. In the following, we discuss the effect of the vortex core position on the four-fold screw-symmetry-protected GVL modes.

In Fig.~\ref{fig:tbvm_even}, we show the vortex bound states of the tight-binding Hamiltonian~(\ref{eq:tbmodel}), where the lattice size is chosen to be an even integer ($L=22$), and the vortex core is located at the center of  a plaquette. Figure~\ref{fig:tbvm_even} (a) shows the existence of zero-energy modes with four-fold degeneracy at $k_z =\pm \pi$, whereas the gapless modes disappear in Fig.~\ref{fig:tbvm_even} (b) as $\mu$ is changed. This is in sharp contrast to Fig.~\ref{fig:tbvm} (f).

Comparing Fig.~\ref{fig:tbvm_even} (a) with Fig.~\ref{fig:tbvm} (b), we notice that the gapless modes around $k_z=\pm \pi$ have different eigenvalues of $\widetilde{\mathcal{S}}_{4z}^{\pi}$.
In Fig.~\ref{fig:tbvm} (b), the two gapless modes with the screw eigenvalue $ie^{-i\frac{\pi}{2}}=1$ (blue) have positive velocities ($dE/dk_z$) at $k_z=-\pi$, while their velocities have opposite signs in Fig.~\ref{fig:tbvm_even} (a). The modes with the $\widetilde{\mathcal{S}}_{4z}^\pi$ eigenvalue $-ie^{-i\frac{\pi}{2}}=-1$ (red) are related to those with $+1$ (blue) by the PH symmetry. According to the low-energy effective Hamiltonian (\ref{eq:screw_H2}), the former situation ($v_1^2>v_2^2+v_3^2$) has stable gapless modes (against the perturbation $m_0$) [Fig.~\ref{fig:tbvm_even} (c)], while the latter ($v_1^2<v_2^2+v_3^2$) does not [Fig.~\ref{fig:tbvm_even} (d)].

In the following, we explain the disappearance of the gapless modes using the low-energy Dirac Hamiltonians discussed in Appendix~\ref{app:low-h}.
We note that the $\tilde{\mathcal{S}}_4^{\pi}$-eigenvalues of zero-energy modes depend on the location of the vortex core.
To see this, we consider a 2D square lattice, where the position of each site is defined by $\bm{r}_{(m,n)} = m \bm{e}_x + n \bm{e}_y $ ($m,n \in \mathbb{Z}$). $\bm{e}_x$ and $\bm{e}_y$ are the unit vector in the $x$ and $y$ directions. We set a vortex core at (i) $(m,n)=(0,0)$ or (ii) $(\frac{1}{2},\frac{1}{2})$, which correspond to the site at the origin and the center of a plaquette, respectively. Then, the four-fold rotation around the vortex core transforms the position of sites as $ \bm{r}_{(m,n)} \to \bm{r}_{(-n,m)}$ for (i) and $ \bm{r}_{(m,n)} \to \bm{r}_{(-n+1,m)}$ for (ii). To evaluate this transformation in the momentum space, we consider the phase part of the Bloch wave function $ e^{-i\bm{k} \cdot \bm{r}_{(m,n)}}$. In particular, we focus on the Bloch wave function at the Dirac cones, which are located at the four-fold rotation invariant momentum $\bm{k}_0=(0,0)$ and $\bm{k}_{\pi}=(\pi,\pi)$. In this case, the effect of four-fold rotation is calculated as, for (i),
\begin{align}
 &e^{-i \bm{k}_0 \cdot \bm{r}_{(m,n)}}=1 \to  e^{-i \bm{k}_0 \cdot \bm{r}_{(-n,m)}}=1, \\
 &e^{-i \bm{k}_{\pi} \cdot \bm{r}_{(m,n)}}=1 \to  e^{-i \bm{k}_{\pi} \cdot\bm{r}_{(-n,m)}}=1,
\end{align}
while for (ii), they change to
\begin{align}
 &e^{-i \bm{k}_0 \cdot \bm{r}_{(m,n)}}=1 \to  e^{-i \bm{k}_0 \cdot \bm{r}_{(-n+1,m)}}=1, \\
 &e^{-i \bm{k}_{\pi} \cdot \bm{r}_{(m,n)} }=1 \to  e^{-i \bm{k}_{\pi} \cdot \bm{r}_{(-n+1,m)}}=-1.
\end{align}
Thus, an extra factor $-1$ is multiplied to the four-fold rotation operator $\widetilde{\mathcal{S}}_{4z}^{\pi}$ at $\bm{k}=\bm{k}_{\pi}$ (i.e., $\nu=-1$) when the vortex core is positioned at the center of the plaquette.
As a consequence, the energy and screw operators in the basis of $( | \Psi_{\nu,1}\rangle,| \Psi_{\nu,2}\rangle)$ are changed to
\begin{align}
 &\langle U \tilde{H}_\nu(\bm{k}) U^{\dagger}\rangle = -\frac{\nu c_\nu t_z}{2} \tilde{k}_z
  \begin{pmatrix} 1 & 0 \\ 0 & -1 \end{pmatrix}, \\
 &\langle U \tilde{\mathcal{S}}_{4z}^{\pi} U^{\dagger} \rangle = + \begin{pmatrix} 1 & 0 \\ 0 & -1 \end{pmatrix}, \label{eq:exp_s4z}
\end{align}
where Eq.~(\ref{eq:exp_s4z}) does not contain $\nu$ unlike Eq.~(\ref{eq:ene-eigenv1-2}).
The results are consistent with Fig.~\ref{fig:tbvm_even} (a) around $k_z=\pi$ and the left panel of Fig.~\ref{fig:tbvm_even} (d).

\bibliography{gapless_vortex}

\end{document}